\global\documentclass[pra,floatfix,nofootinbib,superscriptaddress,notitlepage, twocolumn]{revtex4-1} 
\usepackage{amsmath,amssymb,amsfonts}
\usepackage[T1]{fontenc}
\usepackage[latin9]{inputenc}
\usepackage{lmodern} 
\usepackage{color}
\usepackage{bbold}
\usepackage{dsfont}
\usepackage{graphicx}
\usepackage[caption=false]{subfig}
\usepackage[colorlinks]{hyperref}
\usepackage[bottom]{footmisc}
\usepackage{enumerate}
\usepackage{float}
\usepackage{mathtools}
\usepackage{leftindex}
\usepackage{array}
\usepackage{makecell}
\hypersetup{
	colorlinks=true,
	linkcolor=blue,
	filecolor=magenta,      
	urlcolor=blue,
	citecolor=blue
}

\usepackage{empheq}
\usepackage{tikz}
\usepackage[normalem]{ulem}
\DeclareMathAlphabet\mbc{OMS}{cmsy}{b}{n}
\usepackage{balance}

\begin{document}

\global\long\def\eqn#1{\begin{align}#1\end{align}}
\global\long\def\vec#1{\overrightarrow{#1}}
\global\long\def\ket#1{\left|#1\right\rangle }
\global\long\def\bra#1{\left\langle #1\right|}
\global\long\def\bkt#1{\left(#1\right)}
\global\long\def\sbkt#1{\left[#1\right]}
\global\long\def\cbkt#1{\left\{#1\right\}}
\global\long\def\abs#1{\left\vert#1\right\vert}
\global\long\def\cev#1{\overleftarrow{#1}}
\global\long\def\der#1#2{\frac{{d}#1}{{d}#2}}
\global\long\def\pard#1#2{\frac{{\partial}#1}{{\partial}#2}}
\global\long\def\re{\mathrm{Re}}
\global\long\def\im{\mathrm{Im}}
\global\long\def\dd{\mathrm{d}}
\global\long\def\ddd{\mathcal{D}}
\global\long\def\hmb#1{\hat{\mathbf #1}}
\global\long\def\avg#1{\left\langle #1 \right\rangle}
\global\long\def\mr#1{\mathrm{#1}}
\global\long\def\mb#1{{\mathbf #1}}
\global\long\def\mc#1{\mathcal{#1}}
\global\long\def\tr{\mathrm{Tr}}
\global\long\def\dbar#1{\Bar{\Bar{#1}}}
\newcommand\dover[1]{%
\tikz[baseline=(nodeAnchor.base)]{
    \node[inner sep=0] (nodeAnchor) {$#1$}; 
    \draw[line width=0.1ex,line cap=round] 
        ($(nodeAnchor.north west)+(0.0em,0.2ex)$) 
            --
        ($(nodeAnchor.north east)+(0.0em,0.2ex)$) 
        ($(nodeAnchor.north west)+(0.0em,0.5ex)$) 
            --
        ($(nodeAnchor.north east)+(0.0em,0.5ex)$) 
    ;
}}
\global\long\def\nth{$n^{\mathrm{th}}$\,}
\global\long\def\mth{$m^{\mathrm{th}}$\,}
\global\long\def\non{\nonumber}

\newcommand{\orange}[1]{{\color{orange} {#1}}}
\newcommand{\cyan}[1]{{\color{cyan} {#1}}}
\newcommand{\blue}[1]{{\color{blue} {#1}}}
\newcommand{\yellow}[1]{{\color{yellow} {#1}}}
\newcommand{\green}[1]{{\color{green} {#1}}}
\newcommand{\red}[1]{{\color{red} {#1}}}
\newcommand{\ks}[1]{{\color{orange}[KS: {#1}]}}
\newcommand{\ad}[1]{{\color{red}[AD: {#1}]}}
\newcommand{\imgstrut}{\rule{0pt}{0.15\textwidth}}
\global\long\def\todo#1{\orange{{$\bigstar$ \cyan{\bf\sc #1}}$\bigstar$} }

\title{Casimir-Polder potential on an excited atom near an atomic array}

\author{Annyun Das}
\email{annyun@arizona.edu}
\affiliation{Wyant College of Optical Sciences and Department of Physics, University of Arizona, Tucson, AZ 85721}

\author{Kanu Sinha}
\email{kanu@arizona.edu}
\affiliation{Wyant College of Optical Sciences and Department of Physics, University of Arizona, Tucson, AZ 85721}

\begin{abstract}

We develop a microscopic description of the fluctuation-mediated Casimir-Polder (CP) shifts on a `test' two-level atom placed near a two-dimensional  atomic array of two-level atoms.  We derive the resonant and off-resonant CP potentials experienced by the excited test atom using fourth-order perturbation theory, under the assumption that the test atom resonance is far detuned from those of the array atoms. The total potential on the test atom can be described as the sum of the pairwise resonant and off-resonant potentials resulting from its interaction with the individual atoms of the array. We analyze the asymptotic scaling of CP shifts as a function of the test atom-array separation, and its dependence on various system parameters: array spacing and size, and dipole orientation of the array atoms. 
 Our results bridge the  description of CP potential across two distinct regimes: (i) from a  single-atom limit where we recover the well-known two-atom Van der Waals potential, (ii) to a macroscopic boundary limit, where we  demonstrate new asymptotic scaling laws.  We demonstrate  that these scaling laws can be tuned via the microscopic parameters of the atomic array,  establishing atomically-controlled arrays as a versatile platform for tailoring fluctuation-induced QED phenomena.

 \end{abstract}

\maketitle

\section{Introduction}

 Modification of the quantized electromagnetic field in presence of boundaries is a fundamental feature and a powerful tool in quantum electrodynamics (QED), underlying a range of areas from cavity QED~\cite{Haroche} to nanophotonics~\cite{NovotnyBook} to Casimir physics~\cite{Milonni}. For example, by shaping the electromagnetic mode structure with mirrors, cavities, or photonic structures, one can engineer the quantum vacuum itself, tailoring the spectral density of quantum fluctuations near boundaries.  This control underlies fluctuation-induced phenomena such as Casimir-Polder (CP) forces and Purcell decay of atoms, which, at the microscopic level, arise from interactions between the fluctuating dipole moments of an atom and those constituting a nearby boundary~\cite{Casimir1948,CP1948, Purcell1995, CookMilonni1987}.  

Typically electromagnetic boundary conditions are engendered by classical macroscopic objects, with fixed optical properties. However, recent experimental advances have led to the realization of atomically thin mirrors with ordered arrays of thousands of atoms, manifesting efficient and tunable light-matter interfaces~\cite{Rui2020, Srakaew2023}.
Such structured arrays of atoms can exhibit highly collimated and coherent scattering of light~\cite{Facchinetti2016, Bettles2016, AsenjoGarcia2017, Shahmoon2017, Javanainen2019, Ruostekoski2023, Ballantine2021, Jenkins2012, Robicheaux2025}, with applications in  photon storage ~\cite{Facchinetti2016, AsenjoGarcia2017, Guimond2019, Eltohfa2025},  coherent transport of excitations~\cite{Holzinger2022,Rubies-Bigorda2022},  topological and nonlinear quantum optics~\cite{ Bettles2017,Perczel2017, Perczel2017b, Moreno-Cardoner2021, Rusconi2021}, near-zero refractive index~\cite{Ruks2025} and   magnetometry~\cite{Facchinetti2016,Facchinetti2018}.  Thus such atomic arrays open a qualitatively new regime of QED wherein the electromagnetic vacuum and concomitant quantum fluctuation phenomena can be  tailored via the microscopic properties of the boundary~\cite{Chang2012, Mirhosseini2019, Das2025}. The electromagnetic boundary conditions are thus dynamically established by and even entangled with the atoms constituting the boundary~\cite{Sinha2025, Bekenstein20}. This raises the question: how are quantum fluctuation phenomena modified when the boundary is controlled at the level of individual atoms?

\begin{figure}[b]
    \centering
    \includegraphics[width=\linewidth]{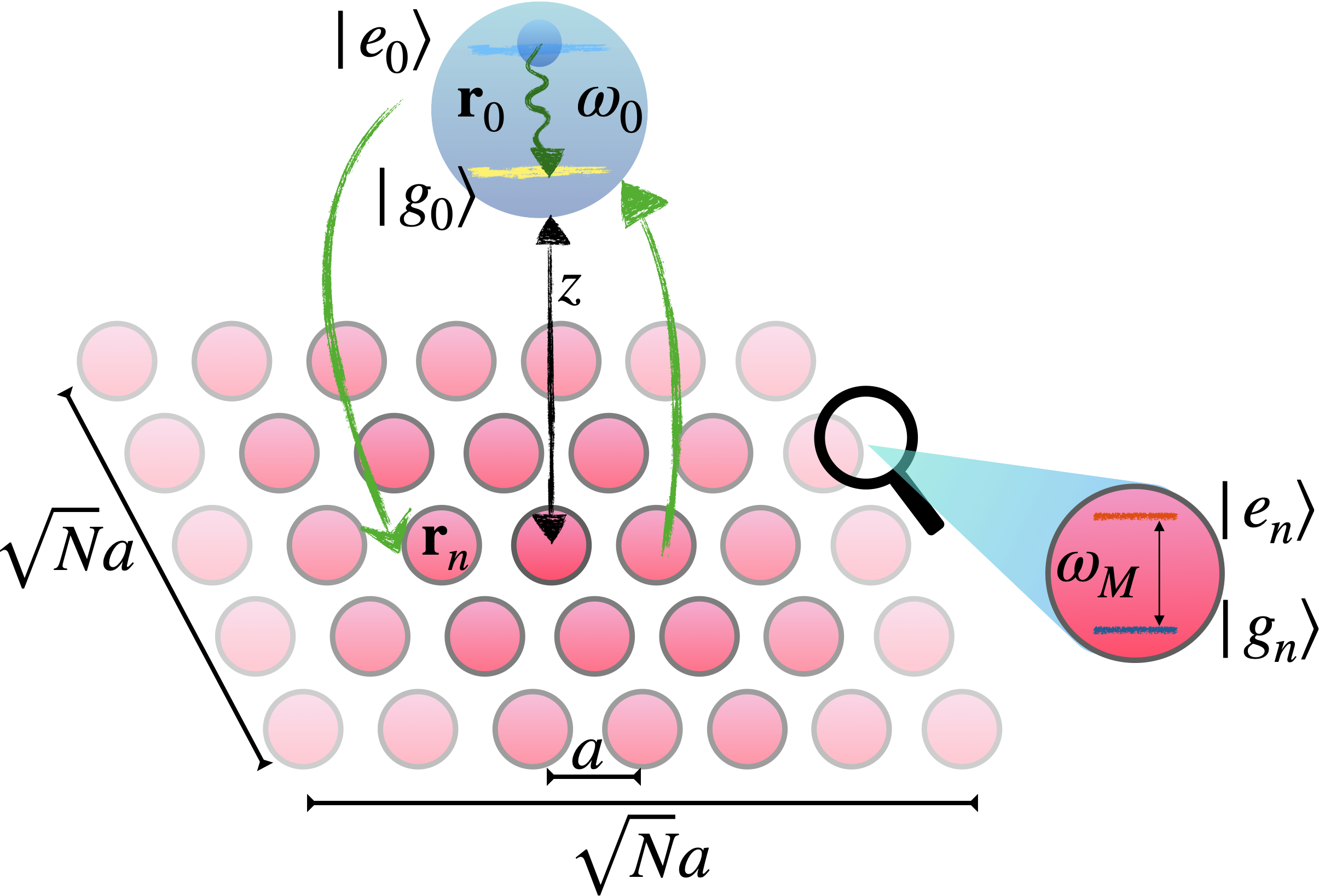}
    \caption{Schematic representation of the model. We consider an excited two-level \textit{test atom} with transition frequency $\omega_0$ placed atop a 2D square lattice of $N$ two-level atoms with transition frequency $\omega_M$ and lattice constant $a$. The array is in the $xy$-plane, with the test atom  placed at a distance $z$ above the central array atom along the $z$-axis.}
    \label{Fig:Schematic}
\end{figure}

From the broader perspective of tailoring quantum fluctuation phenomena, there has been remarkable progress in engineering fluctuation-induced   forces between classical macroscopic bodies~\cite{Gong2021}, e.g.  using geometry~\cite{Chen2002,Levin2010, ARod2008}, nanofabrication~\cite{ARod2011, Intravaia2013}, nonequilibrium environments~\cite{Obrecht2007,Antezza2005, Obrecht2007,Antezza2008, Kruger_2011,Fuchs18a}, and optical and material properties of the interacting bodies or the intervening medium~\cite{Hamaker1937,Hutter1993,Munday2009,Tabor2011,RMPWoods,Jiang2019}. When one of the interacting bodies is microscopic and the other macroscopic, e.g. atoms placed near a  surface, one can leverage the quantum properties of the microscopic object to tailor quantum fluctuation forces.  For example it has been shown that CP forces on atoms and microscopic particles can be engineered by using external drives~\cite{Chang14, Fuchs18a, Fuchs18b, Sinha2020PRA, Jakubec2024},  and by preparing systems in correlated and collective quantum states~\cite{ Behunin10, Fuchs_2018, CollCP18, Jones2018}. Specifically, for excited atoms, the resonant component of the CP potential depends exclusively on the response of the electromagnetic environment at the atomic resonance, enabling one to tailor the CP potential via resonances of nearby media~\cite{Chang14, Goban2014, Laliotis2014, Laliotis2015, LeKien2022, Du2026}. Atomic arrays present a yet unexplored scenario where the medium itself can exhibit quantum behavior and can be manipulated at an atomic scale. This opens a new toolbox  for engineering fluctuation phenomena near \textit{atomically-controlled boundaries}.  Such a boundary, created atom-by-atom, also motivates a crossover question: how does the CP interaction interpolate between the familiar limiting cases of two-atom Van der Waals interaction and that between an atom and a macroscopic medium?

In this work we present a framework to describe the CP potential on a
test atom placed near a  two-dimensional (2D) atomic array. Section~\ref{Sec:model} presents the system setup and Hamiltonian.  In Section~\ref{Sec:CP} we derive the CP potential on the test atom using fourth-order perturbation theory, assuming that the test atom and the array atoms are far detuned from each other. We find that the resonant and off-resonant shifts can be described as a pairwise sum over the individual CP interactions between the test atom and the array atoms. In Section~\ref{Sec:param} we show that the resulting CP potentials can be decomposed into a medium-like `bulk' contribution and a single-atom-like `boundary' part. We analyze the CP potential asymptotic scaling as a function of the test atom-array separation, and study its dependence on various array parameters: array spacing and size, and dipole orientations of the array atoms. We present our conclusions and outlook in Section~\ref{Sec:conclusion}.

\section{Model}
\label{Sec:model}

 We consider a two-dimensional array of $N$ identical ground state two-level atoms with excited and ground states $\ket {e_\mb{n}}$ and $\ket {g_\mb{n}}$ and transition frequency $\omega_M$\footnote{ We assume that $\omega_M$ is the dressed frequency of the array atoms, which includes any vacuum-induced shifts from their mutual interactions. In the limit of an infinite array, which we will assume, such a shift is identical for each atom in the array.} (see Fig.~\ref{Fig:Schematic}). The array atoms, indexed as $\mb{n}\equiv (n_x, n_y)$, with  $ n_{x,y}\in\cbkt{- \sqrt{N}/2,\dots, \sqrt{N}/2}$, are placed on a square lattice with a lattice constant \textit{a} that occupies the $z=0$ plane. Placed at a distance $z$, perpendicularly above the central array atom (placed at $\cbkt{x,y,z}=\cbkt{0,0,0}$), is  an excited two-level \textit{test} atom $\bkt{\cbkt{x,y,z}=\cbkt{0,0,z}}$. The  excited and ground states $\ket {e_0}$ and $\ket {g_0}$ of the test atom are separated by the transition frequency $\omega_0$, such that $\omega_0-\omega_M=\delta$. The total Hamiltonian of the system $\hat H=\hat H_0+\hat H_\text{int}$ consists of the free atoms+field Hamiltonian
\eqn{\label{Eq:FreeHamiltonian}
\hat H_0 = &\hbar\omega_0\hat \sigma_0^+\hat \sigma_0^- +  \sum_{\mb{n}}\hbar\omega_M\hat \sigma_\mb{n}^+\hat \sigma_\mb{n}^-\non\\&+\hbar \sum_\lambda\int d\omega~\omega\int d^3r~\mathbf{\hat f^\dagger}_\lambda\bkt{\textbf{r},\omega}\cdot\mathbf{\hat f}_\lambda\bkt{\textbf{r},\omega}
}
Here, $\hat \sigma_0^{\pm}$  and $\hat \sigma_\mb{n}^{\pm}$ are the ladder operators for the test  atom and the array atom indexed  $ \mb{n} = (n_x, n_y)$, $\cbkt{\mathbf{\hat f}_\lambda^\dagger\bkt{\textbf{r},\omega},\mathbf{\hat f}_\lambda\bkt{\textbf{r},\omega}}$ are the creation and annihilation operators for a photon with energy $\hbar\omega$ at position $\mathbf{r}$, with $ \lambda = e,m$ corresponding to the quantum noise polarization $ (e)$ or magnetization $(m)$~\cite{VogelWelsch, Buhmann1, Buhmann2}. These operators follow the canonical commutation relations $ \sbkt{\hat{ f}_{\lambda,i}^\dagger(\mb{r}, \omega), \hat{ {f}}_{\lambda',j}(\mb{r}', \omega')} = \delta_{\lambda \lambda'} \delta(\omega - \omega')\delta(\mb{r} - \mb{r}')\delta_{ij}$, where $ \cbkt{i, j} \in \cbkt{x,y,z}$ refer to the polarization axis of the field excitation.

$\hat H_\text{int}=-\hat {\mb{d}}_0\cdot\mathbf{\hat E}\bkt{\mathbf{r}_0}- \sum_{\mb{n} }\mathbf{\hat d}_\mb{n}\cdot\mathbf{\hat E}\bkt{\mathbf{r}_\mb{n}}$ is the electric-dipole interaction Hamiltonian, where $\mathbf{\hat d}_\mb{n} = \mathbf d_\mb{n}\bkt{\hat \sigma_\mb{n}^++\hat \sigma_\mb{n}^-}$ is the dipole operator for the array atom indexed $\mb{n}$, located at position $ \mb{r}_\mb{n} = \cbkt{n_xa, n_ya, 0} $, and 
\begin{equation}\label{Eq:ElectricField}
\begin{aligned}
\mathbf{\hat E}\bkt{\mathbf r}=\int d^3r'\int d\omega \sum_\lambda \left[ \dbar{ G}_\lambda(\mb{r}, \mb{r}', \omega)\cdot \hat{\mb{f}}_\lambda\bkt{\mb{r}', \omega} \right.\\ \left.+\hat{\mb{f}}^\dagger_\lambda\bkt{\mb{r}', \omega} \cdot \dbar {G}^\dagger_\lambda(\mb{r}, \mb{r}', \omega)\right]
\end{aligned}
\end{equation}
is the quantized electric field. The coefficients $ \dbar G_\lambda (\mb{r}, \mb{r}' ,\omega)$ are proportional to the free-space Green tensor $\dbar G\bkt{\mb r,\mb r',\omega}$, representing the propagator of the electromagnetic  field from $\mb r'$ to $\mb r$ (see Appendix~\ref{Appendix:Green} for details)~\cite{Buhmann1, VogelWelsch}. More explicitly,

    \begin{widetext}
        
    \eqn{\label{Eq:InteractionHamiltonian}
\hat H_\text{int}=\sum_{\lambda}\int d\omega\int d^3r& \sbkt{\hat \sigma_0^+\textbf{d}_0\cdot \dbar{G}_{\lambda}(\textbf{r}_0,\textbf{r},\omega)\cdot\hat{\mb{f}}_{\lambda}\bkt{\mb{r}, \omega}+ \hat \sigma_0^+ {\hat{\mb{f}}^\dagger_{\lambda}\bkt{\mb{r}, \omega}\cdot\dbar{G}^\dagger_{\lambda}(\textbf{r}_0,\textbf{r},\omega)}\cdot\textbf{d}_0 \right.\non\\
& \left.+ \sum_\mb{n} \hat \sigma_\mb{n}^+\textbf{d}_\mb{n}\cdot \dbar{G}_{\lambda}(\textbf{r}_\mb{n},\textbf{r},\omega)\cdot\hat{\mb{f}}_{\lambda}\bkt{\mb{r}, \omega}+ \hat \sigma_\mb{n}^+ {\hat{\mb{f}}^\dagger_{\lambda}\bkt{\mb{r}, \omega}\cdot\dbar{G}^\dagger_{\lambda}(\textbf{r}_\mb{n},\textbf{r},\omega)}\cdot\textbf{d}_\mb{n}+ \text{H.c.}}
}
\end{widetext}
where the first and second terms denote the co- and counter-rotating terms of the interaction between individual atoms and the EM field. 
We note that we have not made a rotating-wave approximation  in the above interaction Hamiltonian, allowing for both  off-resonant and resonant interactions between the atoms and the field. We will now evaluate the CP potential resulting on the test atom from the above Hamiltonian.

\begin{table*}[t]
\begin{center}
     \begin{tabular}{ | c | c| c | c | c |  } 
    \hline
    \makecell{Feynman\\
          diagram}& \makecell{$t_1:$ $ \ket{m^0}$ }& \makecell{$t_2:$ $ \ket{l^0}$ } & \makecell{$t_3:$ $ \ket{k^0}$ } &  $ D_p $ \\
         \hline
          \raisebox{-0.5\height}{\includegraphics[width=0.08\textwidth]{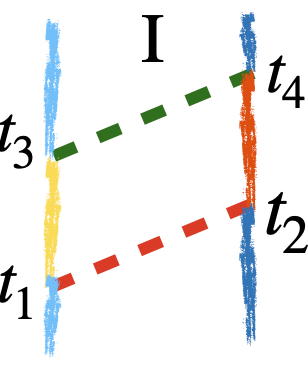}} &  \makecell{$\ket{m^0} = \ket{g_0}\ket{g_\mb{n}} \ket{1_\omega}$\\ 
          \\
          $E_m^0 = \hbar \omega$}& \makecell{$\ket{l^0} = \ket{g_0}\ket{e_\mb{n}} \ket{\cbkt{0}}$\\ 
          \\
          $E_l^0 = \hbar \omega_M$} & \makecell{$\ket{k^0} = \ket{e_0}\ket{e_\mb{n}} \ket{1_{\omega'}}$\\ 
          \\
          $E_k^0 = \hbar (\omega_0 + \omega_M + \omega')$}  &  $  - \bkt{\omega_0 - \omega} \bkt{\omega_0 - \omega_M}\bkt{\omega' + \omega_M}$\\
        \hline
        \raisebox{-0.5\height}{\includegraphics[width=0.08\textwidth]{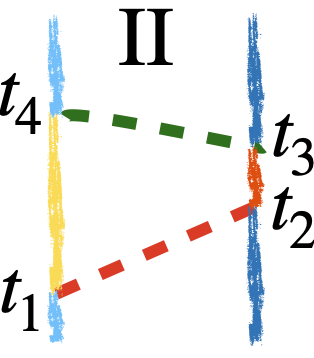}} &  \makecell{$\ket{m^0} = \ket{g_0}\ket{g_\mb{n}} \ket{1_\omega}$\\ 
          \\
          $E_m^0 = \hbar \omega$}& \makecell{$\ket{l^0} = \ket{g_0}\ket{e_\mb{n}} \ket{\cbkt{0}}$\\ 
          \\
          $E_l^0 = \hbar \omega_M$} & \makecell{$\ket{k^0} = \ket{g_0}\ket{g_\mb{n}} \ket{1_{\omega'}}$\\ 
          \\
          $E_k^0 = \hbar \omega'$}  &  $ \bkt{\omega_0 - \omega} \bkt{\omega_0 - \omega_M}\bkt{\omega_0 - \omega'}$\\
          \hline
          \raisebox{-0.5\height}{\includegraphics[width=0.08\textwidth]{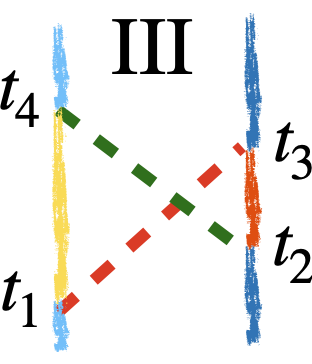}} &  \makecell{$\ket{m^0} = \ket{g_0}\ket{g_\mb{n}} \ket{1_\omega}$\\ 
          \\
          $E_m^0 = \hbar \omega$}& \makecell{$\ket{l^0} = \ket{g_0}\ket{e_\mb{n}} \ket{1_{\omega}}\ket{1_{\omega'}}$\\ 
          \\
          $E_l^0 = \hbar \bkt{\omega_M + \omega + \omega'}$} & \makecell{$\ket{k^0} = \ket{g_0}\ket{g_\mb{n}} \ket{1_{\omega'}}$\\ 
          \\
          $E_k^0 = \hbar \omega'$}  &  $  \bkt{\omega_0 - \omega} \bkt{\omega_0 - \omega_M - \omega - \omega'}\bkt{\omega_0 - \omega'}$\\
          \hline
          \raisebox{-0.5\height}{\includegraphics[width=0.08\textwidth]{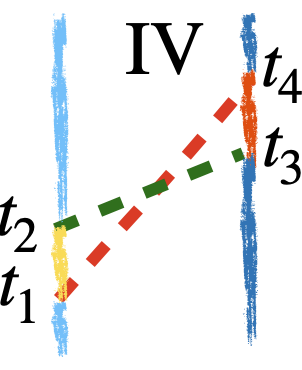}} &  \makecell{$\ket{m^0} = \ket{g_0}\ket{g_\mb{n}} \ket{1_\omega}$\\ 
          \\
          $E_m^0 = \hbar \omega$}& \makecell{$\ket{l^0} = \ket{e_0}\ket{g_\mb{n}} \ket{1_{\omega}}\ket{1_{\omega'}}$\\ 
          \\
          $E_l^0 = \hbar \bkt{\omega_0 + \omega + \omega'}$} & \makecell{$\ket{k^0} = \ket{e_0}\ket{e_\mb{n}} \ket{1_{\omega}}$\\ 
          \\
          $E_k^0 = \hbar \bkt{\omega_0+ \omega_M  + \omega}$}  &  $  \bkt{\omega_0 - \omega} \bkt{\omega + \omega'}\bkt{  \omega_M + \omega}$\\
          \hline
          \raisebox{-0.5\height}{\includegraphics[width=0.08\textwidth]{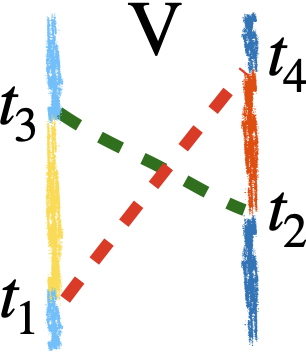}} &  \makecell{$\ket{m^0} = \ket{g_0}\ket{g_\mb{n}} \ket{1_\omega}$\\ 
          \\
          $E_m^0 = \hbar \omega$}& \makecell{$\ket{l^0} = \ket{g_0}\ket{e_\mb{n}} \ket{1_{\omega}}\ket{1_{\omega'}}$\\ 
          \\
          $E_l^0 = \hbar \bkt{\omega_M + \omega + \omega'}$} & \makecell{$\ket{k^0} = \ket{e_0}\ket{e_\mb{n}} \ket{1_{\omega}}$\\ 
          \\
          $E_k^0 = \hbar \bkt{\omega_0+ \omega_M  + \omega}$}  &  $ -\bkt{\omega_0 - \omega} \bkt{\omega_0 -\omega_M - \omega - \omega'}\bkt{  \omega_M + \omega}$\\
          \hline
          \raisebox{-0.5\height}{\includegraphics[width=0.08\textwidth]{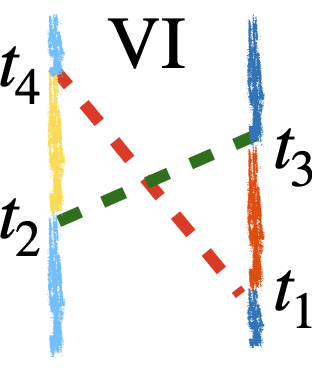}} &  \makecell{$\ket{m^0} = \ket{e_0}\ket{e_\mb{n}} \ket{1_\omega}$\\ 
          \\
          $E_m^0 = \hbar \bkt{\omega_0 + \omega_M + \omega}$}& \makecell{$\ket{l^0} = \ket{g_0}\ket{e_\mb{n}} \ket{1_{\omega}}\ket{1_{\omega'}}$\\ 
          \\
          $E_l^0 = \hbar \bkt{\omega_M + \omega + \omega'}$} & \makecell{$\ket{k^0} = \ket{g_0}\ket{g_\mb{n}} \ket{1_{\omega}}$\\ 
          \\
          $E_k^0 = \hbar \omega$}  &  $  -\bkt{\omega_M + \omega} \bkt{\omega_0 -\omega_M - \omega - \omega'}\bkt{  \omega_0 - \omega}$\\
          \hline
          \raisebox{-0.5\height}{\includegraphics[width=0.08\textwidth]{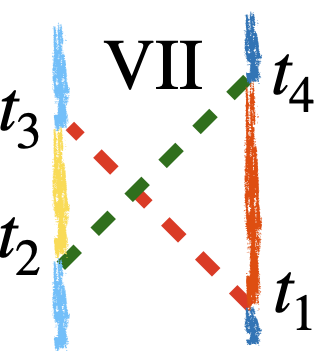}} &  \makecell{$\ket{m^0} = \ket{e_0}\ket{e_\mb{n}} \ket{1_\omega}$\\ 
          \\
          $E_m^0 = \hbar \bkt{\omega_0 + \omega_M + \omega}$}& \makecell{$\ket{l^0} = \ket{g_0}\ket{e_\mb{n}} \ket{1_{\omega}}\ket{1_{\omega'}}$\\ 
          \\
          $E_l^0 = \hbar \bkt{\omega_M + \omega + \omega'}$} & \makecell{$\ket{k^0} = \ket{e_0}\ket{e_\mb{n}} \ket{1_{\omega'}}$\\ 
          \\
          $E_k^0 = \hbar \bkt{\omega_0 + \omega_M + \omega'}$}  &  $   \bkt{\omega_M + \omega} \bkt{\omega_0 -\omega_M - \omega - \omega'}\bkt{  \omega_M + \omega'}$\\
          \hline
          \raisebox{-0.5\height}{\includegraphics[width=0.08\textwidth]{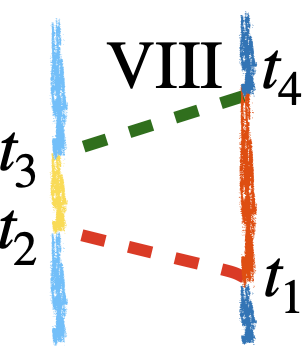}} &  \makecell{$\ket{m^0} = \ket{e_0}\ket{e_\mb{n}} \ket{1_\omega}$\\ 
          \\
          $E_m^0 = \hbar \bkt{\omega_0 + \omega_M + \omega}$}& \makecell{$\ket{l^0} = \ket{g_0}\ket{e_\mb{n}} \ket{\cbkt{0}}$\\ 
          \\
          $E_l^0 = \hbar \omega_M $} & \makecell{$\ket{k^0} = \ket{e_0}\ket{e_\mb{n}} \ket{1_{\omega'}}$\\ 
          \\
          $E_k^0 = \hbar \bkt{\omega_0 + \omega_M + \omega'}$}  &  $  \bkt{\omega_M + \omega} \bkt{\omega_0 -\omega_M }\bkt{  \omega_M + \omega'}$\\
          \hline
          \raisebox{-0.5\height}{\includegraphics[width=0.08\textwidth]{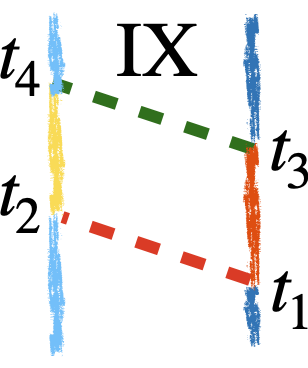}} &  \makecell{$\ket{m^0} = \ket{e_0}\ket{e_\mb{n}} \ket{1_\omega}$\\ 
          \\
          $E_m^0 = \hbar \bkt{\omega_0 + \omega_M + \omega}$}& \makecell{$\ket{l^0} = \ket{g_0}\ket{e_\mb{n}} \ket{\cbkt{0}}$\\ 
          \\
          $E_l^0 = \hbar \omega_M $} & \makecell{$\ket{k^0} = \ket{g_0}\ket{g_\mb{n}} \ket{1_{\omega'}}$\\ 
          \\
          $E_k^0 = \hbar \omega'$}  &  $  -\bkt{\omega_M + \omega} \bkt{\omega_0 -\omega_M }\bkt{  \omega_0 - \omega'}$\\
          \hline
          \raisebox{-0.5\height}{\includegraphics[width=0.08\textwidth]{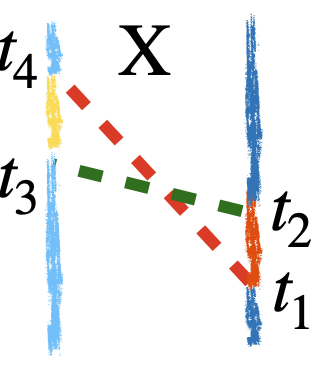}} &  \makecell{$\ket{m^0} = \ket{e_0}\ket{e_\mb{n}} \ket{1_\omega}$\\ 
          \\
          $E_m^0 = \hbar \bkt{\omega_0 + \omega_M + \omega}$}& \makecell{$\ket{l^0} = \ket{e_0}\ket{g_\mb{n}} \ket{1_\omega} \ket{1_{\omega'}}$\\ 
          \\
          $E_l^0 = \hbar \bkt{\omega_0 + \omega + \omega'} $} & \makecell{$\ket{k^0} = \ket{g_0}\ket{g_\mb{n}} \ket{1_{\omega}}$\\ 
          \\
          $E_k^0 = \hbar \omega$}  &  $  \bkt{\omega_M + \omega} \bkt{\omega + \omega' }\bkt{  \omega_0 - \omega}$\\
          \hline
          \raisebox{-0.5\height}{\includegraphics[width=0.08\textwidth]{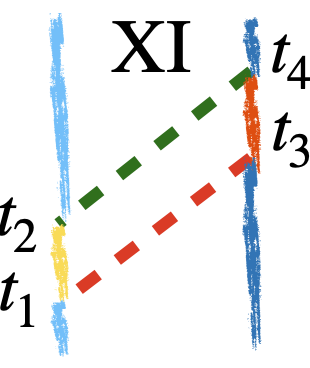}} &  \makecell{$\ket{m^0} = \ket{g_0}\ket{g_\mb{n}} \ket{1_\omega}$\\ 
          \\
          $E_m^0 = \hbar \omega$}& \makecell{$\ket{l^0} = \ket{e_0}\ket{g_\mb{n}} \ket{1_\omega} \ket{1_{\omega'}}$\\ 
          \\
          $E_l^0 = \hbar \bkt{\omega_0 + \omega + \omega'} $} & \makecell{$\ket{k^0} = \ket{e_0}\ket{e_\mb{n}} \ket{1_{\omega'}}$\\ 
          \\
          $E_k^0 = \hbar \bkt{\omega_0 + \omega_M + \omega'}$}  &  $  \bkt{ \omega_M +  \omega'} \bkt{\omega + \omega' }\bkt{  \omega_0 - \omega}$\\
          \hline
          \raisebox{-0.5\height}{\includegraphics[width=0.08\textwidth]{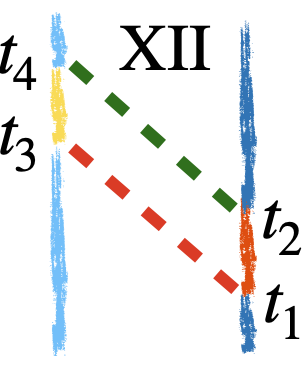}} &  \makecell{$\ket{m^0} = \ket{e_0}\ket{e_\mb{n}} \ket{1_\omega}$\\ 
          \\
          $E_m^0 = \hbar \bkt{\omega_0 + \omega_M + \omega}$}& \makecell{$\ket{l^0} = \ket{e_0}\ket{g_\mb{n}} \ket{1_\omega} \ket{1_{\omega'}}$\\ 
          \\
          $E_l^0 = \hbar \bkt{\omega_0 + \omega + \omega'} $} & \makecell{$\ket{k^0} = \ket{g_0}\ket{g_\mb{n}} \ket{1_{\omega'}}$\\ 
          \\
          $E_k^0 = \hbar \omega'$}  & $ \bkt{ \omega_M +  \omega} \bkt{\omega + \omega' }\bkt{  \omega_0 - \omega'}$  \\
          \hline
    \end{tabular}
    \caption{ The 12 processes that contribute to the CP potential seen by the test atom with virtual states $ \cbkt{\ket{m^0}, \ket{l^0}, \ket{k^0}}$ (Eq.~\eqref{Eq:FourthOrderPerturbation} ) and the denominators $ D_p$ (Eq.~\eqref{Eq:dp}). Time increases vertically with $t_1<t_2<t_3<t_4$ in increasing order. The internal excited and ground states of the test atom are denoted by light blue $(\ket{e_0})$ and yellow $(\ket{g_0})$; those of the array atoms are denoted by red $(\ket{e_\mb{n}})$ and dark blue $(\ket{g_\mb{n}})$, respectively.  The red-dashed (green-dashed) lines refer to the first (second) released virtual photon with frequency $\omega$ ($\omega'$).   }\label{Table:dp}
    \end{center}
\end{table*}


\section{Casimir-Polder potential of the test atom}
\label{Sec:CP}

We compute the energy-shifts of the test atom in the presence of the array via perturbation theory, assuming weak atom-field coupling.  To this end, we require fourth-order processes that capture the interaction between the test atom and individual atoms of  the atomic array.  We express the fourth-order energy shift to the test atom as \cite{McLone1965,Buhmann1}:

\eqn{\label{Eq:FourthOrderPerturbation}
\Delta_\alpha^{(4)}=\sum_{k\neq \alpha}\sum_{l\neq \alpha}\sum_{m\neq \alpha}\frac{\bra{\alpha^0}\hat H_\text{int}\ket{k^0}}{E_\alpha^0-E_k^0}\frac{\bra{k^0}\hat H_\text{int}\ket{l^0}}{E_\alpha^0-E_l^0}&\non\\
\frac{\bra{l^0}\hat H_\text{int}\ket{m^0}}{E_\alpha^0-E_m^0}\bra{m^0}\hat H_\text{int}\ket{\alpha^0}&
}
where, $\ket{\alpha^0}=\ket{e_0}\otimes\cbkt{\ket{g_\mb{n}}}\otimes \ket{\cbkt{0}}$  represents the unperturbed eigenket  of $\hat H_0$, for which  we wish to determine the energy shift,  with $ \ket{\cbkt{g_\mb{n}}}$ corresponding to all the array atoms being in the ground state, and $ \ket{\cbkt{0}}$ as the vacuum state of the EM field.  $ E_\alpha^0$ is the energy eigenvalue for the unperturbed state $ \ket{\alpha^0}$. The states $ \cbkt{\ket{k^0}, \ket{l^0}, \ket{m^0}}$ correspond to intermediate states with eigenenergies $ \cbkt{E_k^0, E_l^0, E_m^0}$, respectively.

The  detuning $\delta=\omega_0 - \omega_M $ between the test atom and the array atoms makes the state $\ket{\alpha^0}$ non-degenerate with respect to the unperturbed states in the single-excitation subspace for the array atoms. This allows us to use  non-degenerate perturbation theory to compute the energy shift for the test atom, as given by Eq.~\eqref{Eq:FourthOrderPerturbation}. Furthermore, the large detuning $ \delta$ between the test atom and the array atoms allows one to ignore processes wherein a photon may resonantly scatter multiple times between the array atoms, unlike in the case of a resonant excitation in a 2D atomic array~\cite{Facchinetti2016,Bettles2016,Shahmoon2017, AsenjoGarcia2017}.

Substituting the interaction Hamiltonian in Eq.~\eqref{Eq:InteractionHamiltonian} into Eq.~\eqref{Eq:FourthOrderPerturbation}, we end up with the CP potential (see Appendix~\ref{App:4CP} for details)
\eqn{\label{Eq:FourthOrderShift}
 &U_\mr{CP}=-\frac{\mu_0^2}{\hbar\pi^2}\non\\&\sum_{p=\mr{I}}^{\mr{XII}}\sum_{\mb{n}}\int_0^\infty d\omega \omega^2\int_0^\infty d\omega' \omega'^2 \frac{\im  g_{0\mb{n}}(\omega)\im  g_{\mb{n}0}(\omega') }{D_p},
}
where $  g_{0\mb{n}}(\omega_0)=\mb{d}_0\cdot\dbar G\bkt{\mb r_0,\mb r_\mb{n},\omega_0}\cdot \mb d_\mb{n} $. The above CP potential  consists of twelve processes as listed in Table~\ref{Table:dp};  these capture the fourth-order perturbative shift of the test atom's energy \cite{McLone1965,Power1995,Donaire2015}. Each process diagram starts with an excited test atom $\bkt{\ket {e_0}}$, all ground state array atoms $\bkt{\ket{ g_\mb{n}}}$ and the vacuum field $\ket{\cbkt{0}}$, and comprises of all the processes captured by both RWA and non-RWA terms of the Hamiltonian in Eq. \eqref{Eq:InteractionHamiltonian}.
The denominators 
\eqn{\label{Eq:dp}
D_p \equiv  \frac{1}{\hbar^3}\bkt{E_\alpha^{0} - E_m^{0}} \bkt{E_\alpha^{0} - E_l^{0}}\bkt{E_\alpha^{0} - E_k^{0}}
}
for each process $p = \cbkt{\mr{I},\dots\mr{XII}}$  are listed in Table \ref{Table:dp}. We note that the self-energy corrections to any of the atoms, arising from second-order perturbation do not contribute to the CP potential of the test atom~\cite{Buhmann1}.

Performing the sum over processes $(p)$ in Eq.~\eqref{Eq:FourthOrderShift} yields the CP  potential of the test atom as $U_\mr{CP} =\hbar\bkt{ \Delta \omega^\mr{R} + \Delta \omega^\mr{OR} }$, where (see Appendix \ref{Appendix: Feynman Diagrams} for details):

\eqn{\label{Eq:omegaR}
\Delta \omega^\mr{R} = & \frac{2\mu_0^2}{\hbar^2}\frac{\omega_0^4\omega_M }{\delta\bkt{\omega_0+\omega_M} }\sum_{\mb{n}}\re\sbkt{g_{0\mb{n}}(\omega_0)g_{\mb{n}0}(\omega_0)}\\
\label{Eq:omegaOR}
\Delta \omega^\mr{OR} = & \frac{2\mu_0^2\omega_0\omega_M}{\hbar^2\pi}\sum_{\mb{n}}\int_0^\infty d\xi\xi^4\frac{ g_{0\mb{n}}(i\xi)g_{\mb{n}0}(i\xi)}{\bkt{\xi^2+\omega_0^2}\bkt{\xi^2+\omega_M^2}}.
}
 Here $ \Delta \omega^\mr{R}$  corresponds to the resonant CP  shift, resulting from the interaction  mediated via a resonantly emitted photon by the test atom at frequency $ \omega_0$. We would like to note that `resonant' here does not imply a resonance between the test atom and the array atoms, but rather to the contribution to the total CP shift of the test atom that arises from on-shell processes. $ \Delta \omega^\mr{OR}$ represents the off-resonant  CP shift arising from the interaction mediated between the test atom and the array via virtual (off-resonant) photons. The off-resonant  CP shift for a ground state atom near an atom array was recently studied in~\cite{Ye2026}. The  shift  $ \Delta \omega^\mr{OR}$ in Eq.~\eqref{Eq:omegaOR} agrees with the general form of the off-resonant shift in \cite{Ye2026} , accounting for the differences in the two models. Compared to the CP potential of two ground state atoms placed near each other,  the off-resonant CP potential between an excited test atom and a ground-state atom  has an opposite sign~\cite{Buhmann1, Safari2015}. We further remark that the above expressions for the resonant and off resonant CP shifts apply for any geometric arrangement of the array atoms, including non-planar and non-periodic geometries.
 
  The factor of  $1/\delta$ in the resonant CP shift  (Eq.~\eqref{Eq:ResonantNonRetardedZZ}) implies that the resonant CP shift can be enhanced for small detunings. However, for $\delta\rightarrow0$,  our analysis, carried out under the non-degenerate perturbation scheme, would be invalid. As before, for sufficiently large detunings we do not need to consider multiple scattering processes of a resonant photon within the 2D array; thus the present treatment within fourth-order perturbation theory is valid only for a far-detuned test atom.

\section{Asymptotic scaling laws and dependence on array parameters}
\label{Sec:param}

\subsection{Crossover from single-atom to macroscopic medium limit}

We now apply the general expressions for the CP potentials, derived in Eqs.~\eqref{Eq:omegaR} and \eqref{Eq:omegaOR} to the specific case of a 2D atomic array, focusing, in particular, on the scaling of the CP potential with test atom-array separation $ z$. Understanding such scaling laws for the case of two particles has been historically instrumental in the development of the QED description of fluctuation forces~\cite{CP1948, Milonni}. Given the microscopic nature of atomic arrays, one can tune various parameters to control the CP potential near such boundaries and the corresponding  asymptotic scaling laws. We specifically focus on three parameters: (1) the lattice spacing $a$, (2) size of the array $N$, and (3)  the orientation of the array dipoles.

 In order to simplify the discrete sum over the array atoms in the CP potentials (Eqs.\eqref{Eq:omegaR} and \eqref{Eq:omegaOR}), we assume a large number of atoms in the array ($N\rightarrow\infty$) and that all the array atomic dipoles are identical ($ \mb{d}_\mb{n}  \equiv \mb{d}$ for all $\mb{n}$). We use a two-dimensional extension of the Euler-Maclaurin formula to decompose the sum over lattice points as follows~\cite{stegun, Guo_2021}  (see Appendix~\ref{App:EM} for details) 
 \begin{widetext}
\eqn{\label{Eq:EulerMaclaurinExpansion}
&\lim_{N\to\infty}\sum_{n_x=-\sqrt{N}/2}^{\sqrt{N}/2}\sum_{n_y=-\sqrt{N}/2}^{\sqrt{N}/2}f\bkt{n_x,n_y}=  \underbrace{\frac{4}{a^2}\int_{0}^{\sqrt{N}a/2} dy{{\int_{0}^{\sqrt{N}a/2}dx~f\bkt{x/a,y/a;\omega}} }}_\text{`Bulk'/medium-like contribution} \non\\
& +\underbrace{\frac{2}{a}\sbkt{\int_{0}^{\sqrt{N}a/2} dy f\bkt{0,y/a;\omega}
+\int_{0}^{\sqrt{N}a/2}dxf\bkt{x/a,0; \omega} }}_\text{`Edge' contribution}
 +\underbrace{f\bkt{0,0; \omega}}_\text{`Vertex'/single-atom-like contribution}, 
}
\end{widetext}
where $ f(\mb n;\omega)\equiv g_{0\mb n} (\omega)g_{\mb n0}(\omega) $. This decomposition allows us to see a crossover from a `medium-like' to `single-atom-like' limit of the CP interaction between the test atom and the array.  Here the first integral term in Eq.~\eqref{Eq:EulerMaclaurinExpansion} represents the contribution from the `bulk' part of the 2D lattice which captures the  response of the dense lattice; while the `vertex' term recovers the single-atom limit. Particularly, we note that the first term has a factor of $ 1/a^2$, corresponding to the effective scattering cross-section of the array,  which enhances the bulk collective response  for dense arrays~\cite{Bettles2016, Shahmoon2017}. The vertex term  recovers the `single-atom-like' limit  for when the lattice spacing $a\gg\lambda_0, z$  (where $ \lambda_0  = 2\pi c/\omega_0$ is the resonant wavelength of the test atom). This corresponds to the scenario where the test atom can resolve individual atoms of a sparse array. The edge term that scales with $ \sim 1/a$ contributes in the regime of intermediate array densities. Throughout the analysis below we assume that the lattice spacing $a$ is sufficiently large compared to the atomic size, such that there is no spatial overlap of the individual atomic wavefunctions.

In what follows, we analyze the asymptotic scaling laws for the CP potential seen by the test atom for two specific orientations of the array dipoles: along the $ z$ and $ x$ axes, respectively. We will consider that the test atom's dipole is oriented along the $z$-axis, although the following calculations can easily be extended to other orientations of the test atom dipole.  In deriving the analytical expressions and scalings for the asymptotic limits, we focus on the  medium-like (bulk) and single-atom-like (vertex) contributions (i.e., first and third terms in Eq.~\eqref{Eq:EulerMaclaurinExpansion}, respectively), which become prominent for dense and sparse arrays, respectively.
 
\subsection{Asymptotic CP shifts for $z$-oriented array dipoles}

We simplify the CP potentials (Eqs.~\eqref{Eq:omegaR} and \eqref{Eq:omegaOR} ) by  plugging in  the free space Green tensor (Eq. \eqref{Eq:Green's Tensor Real Space}), and using the Euler-Maclaurin formula in the limit $N\to\infty$ (see Appendix \ref{Appendix: Scaling Laws} for details). We discuss specifically the non-retarded and retarded regimes of the test atom-array interaction corresponding to  $ z/\lambda_0 \ll 1$ and   $z/\lambda_0 \gg 1$, respectively.
\subsubsection{Non-retarded regime $\bkt{ z/\lambda_0\ll1}$}  We have for the resonant frequency shift $\bkt{\Delta\omega^\mr{R}}$ in the non-retarded regime (see Appendix~\ref{App:res} for details): 
    \begin{subequations}
    \label{Eq:ResonantNonRetardedZZ}
    \eqn{\label{Eq:ResonantNonRetardedZZsparse}
\lim_{N\to\infty}\left.\frac{\Delta\omega^{\mr{R}}}{\gamma_0}\right\vert _{\tilde z \ll1; \tilde a \gg1}\approx&\frac{9\gamma_0\omega_M}{2\delta\bkt{\omega_0+\omega_M}}\frac{1}{\tilde z^6},\\
\label{Eq:ResonantNonRetardedZZdense}
\lim_{N\to\infty}\left.\frac{\Delta\omega^{\mr{R}}}{\gamma_0}\right\vert _{\tilde z \ll1; \tilde a \ll1}\approx&
\frac{27\pi\gamma_0\omega_M}{16\delta\bkt{\omega_0+\omega_M}}\frac{1}{\tilde a^2\tilde z^4},
}
    \end{subequations}
   where  $\tilde z\equiv k_0z$, $\tilde a\equiv k_0a$ and $\gamma_0 =\frac{d^2\omega_0^3}{3\pi\epsilon_0\hbar c^3}$ is the free space spontaneous emission rate for the test atom.  In the limit of large lattice spacing $a\gg\lambda_0$, the primary contribution to the resonant CP potential is from Eq.~\eqref{Eq:ResonantNonRetardedZZsparse},  corresponding to the limit where only a single lattice atom interacts with the test atom. In the large lattice-spacing limit, the CP potential goes as  $\sim 1/\tilde z^6$, similar to the typical Van der Waals potential between two atoms~\cite{vdWaals}. Eq.~\eqref{Eq:ResonantNonRetardedZZsparse} for the resonant CP potential in the single-atom limit is consistent with the CP potential between an  excited and a ground state atom, as obtained in Ref.~\cite{Milonni2015} (note that they use Gaussian units). For smaller lattice spacings, it can be seen from Eq.~\eqref{Eq:ResonantNonRetardedZZdense} that the scaling law approaches $\sim 1/\tilde z^{4}$ for dense arrays. This corresponds to a distinct scaling law for CP potential in the non-retarded regime. For reference, the resonant CP potential on an atom near a planar macroscopic medium goes as $ \sim 1/\tilde z^3$ in the non-retarded regime~\cite{CP1948}.    
   \begin{figure}[t]
    \centering
    \includegraphics[width=0.8\linewidth]{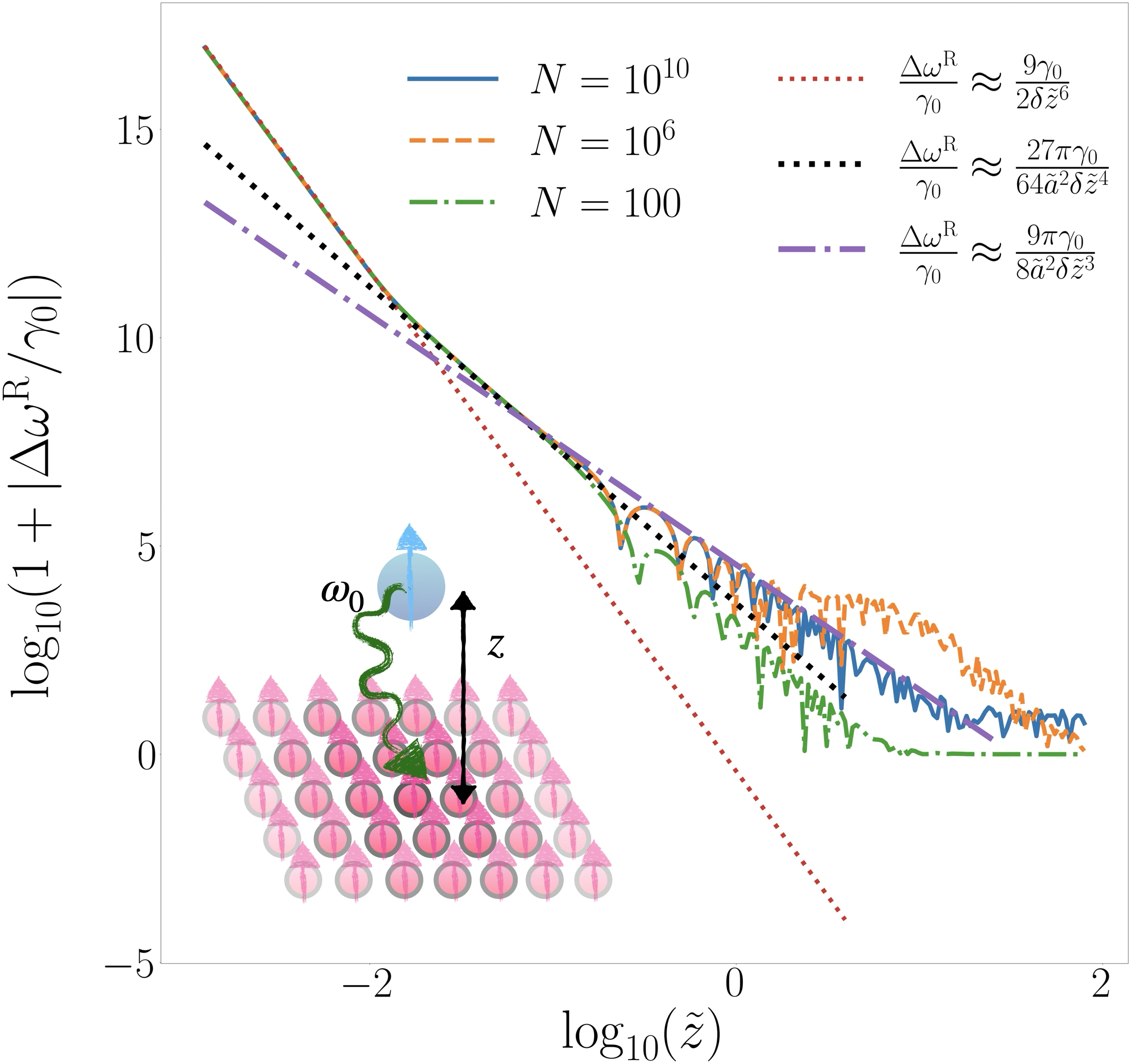}
    \caption{ The resonant shift $\Delta\omega^\mr{R}$ is plotted for $N=100$ (dash-dotted green), $N=10^6$ (dashed orange), and $N=10^{10}$ (solid blue) array atoms, obtained by performing a numerical sum over all lattice points using Eq.~\eqref{Eq:omegaR} for a fixed $\tilde a\approx 10^{-2}$.
    }\label{Fig:CPZZPlots}
\end{figure}

   The off-resonant contribution, $\Delta\omega^\mr{OR}$, in the non-retarded regime can be obtained as (see Appendix~\ref{App:offres} for details):
        \begin{subequations}\label{Eq:OffResonantNonRetardedZZ}        
            \eqn{\label{Eq:OffResonantNonRetardedZZsparse}        \lim_{N\to\infty}\left.\frac{\Delta\omega^\mr{OR}}{\gamma_0}\right\vert_{\tilde z\ll1; \tilde a\gg1}\approx& \frac{9 \gamma_0}{4\bkt{\omega_0+\omega_M}}{\frac{1}{\tilde z^6}},\\
            \label{Eq:OffResonantNonRetardedZZdense} \lim_{N\to\infty}\left.\frac{\Delta\omega^\mr{OR}}{\gamma_0}\right\vert_{\tilde z\ll1; \tilde a \ll1}\approx &\frac{27\pi\gamma_0}{64\bkt{\omega_0+\omega_M}}\frac{1}{\tilde a^2\tilde z^4},
        }
        \end{subequations}
      As seen from the above equations, the off-resonant CP potential, in the large lattice-spacing limit ($a\gg \lambda_0$), falls off as $\sim1/\tilde z^6$; while for subwavelength lattices ($a\ll \lambda_0$), it scales as $\sim1/\tilde z^4$.    Recently, in Ref. \cite{Ye2026}, similar scaling behavior for off-resonant CP potential of an atom near a dense array has been reported for a ground state atom placed near a 2D array with isotropic dipoles, with scaling transition from $1/\tilde z^6\to1/\bkt{\tilde z^4\tilde a^2}$.
            \begin{table}[ht]
\begin{center}
     \begin{tabular}{ | c | c |c| } 
    \hline
         \makecell{CP potential\\ $z$-aligned dipoles} & \makecell{Single-atom limit\\ $a\gg\lambda_0,z$}& \makecell{Dense array limit\\ $a\ll\lambda_0,z$} \\
         \hline
         \raisebox{-0.5\height}{\makecell{\\Non-retarded $\Delta \omega^R$\\
         \includegraphics[width=0.16\textwidth]{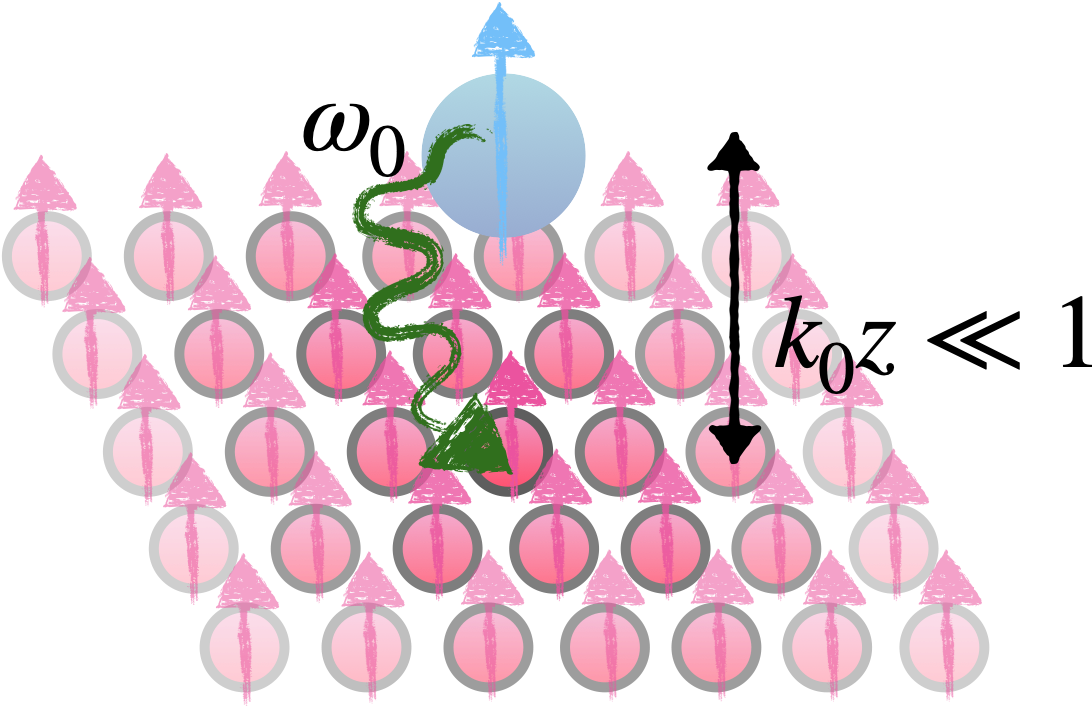}
         \\ Eq.~\eqref{Eq:ResonantNonRetardedZZ}}}
& \raisebox{-3\height}{ \makecell{$1/z^6$ \cite{Milonni2015}}}
& \raisebox{-3\height}{ \makecell{$1/(z^4a^2)$
}} \\
         \hline
         \raisebox{-0.5\height}{\makecell{\\Non-retarded $\Delta\omega^\mr{OR}$\\ \includegraphics[width=0.16\textwidth]{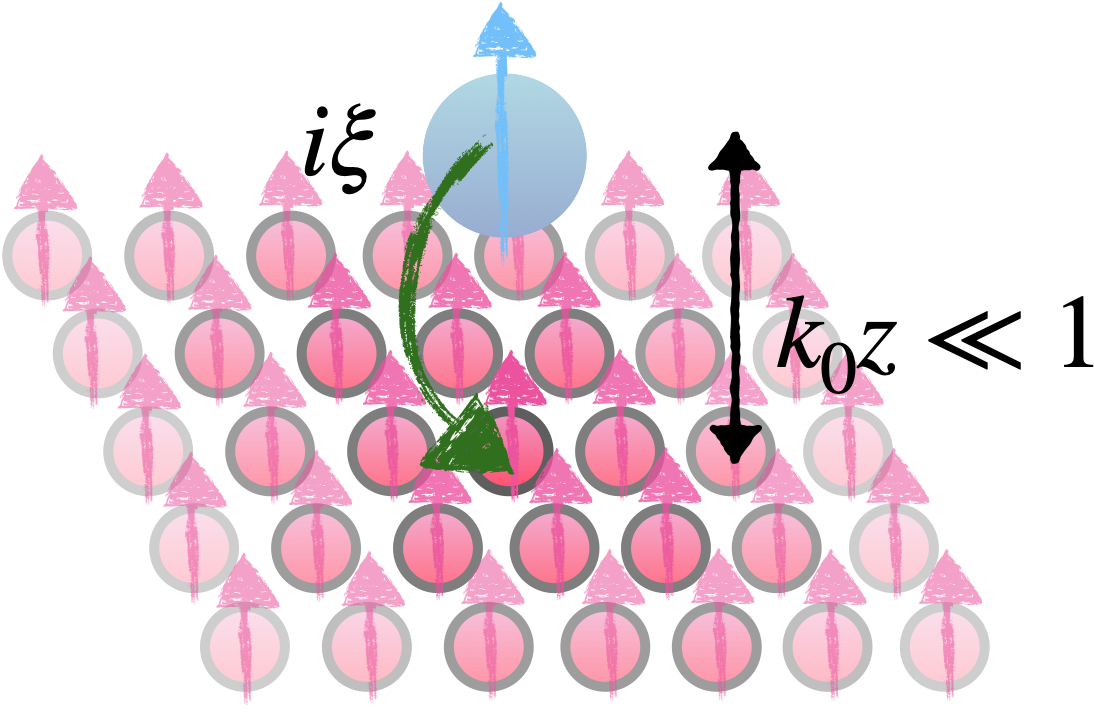}\\
Eq.~\eqref{Eq:OffResonantNonRetardedZZ}}}& \raisebox{-3\height}{ \makecell{$1/z^6$~\cite{vdWaals, CP1948}}} & \raisebox{-3\height}{ \makecell{$1/(z^4a^2)$~\cite{Ye2026}}} \\
         \hline
                \raisebox{-0.5\height}{\makecell{\\Retarded $\Delta \omega^\mr{R}$\\
\includegraphics[width=0.16\textwidth]{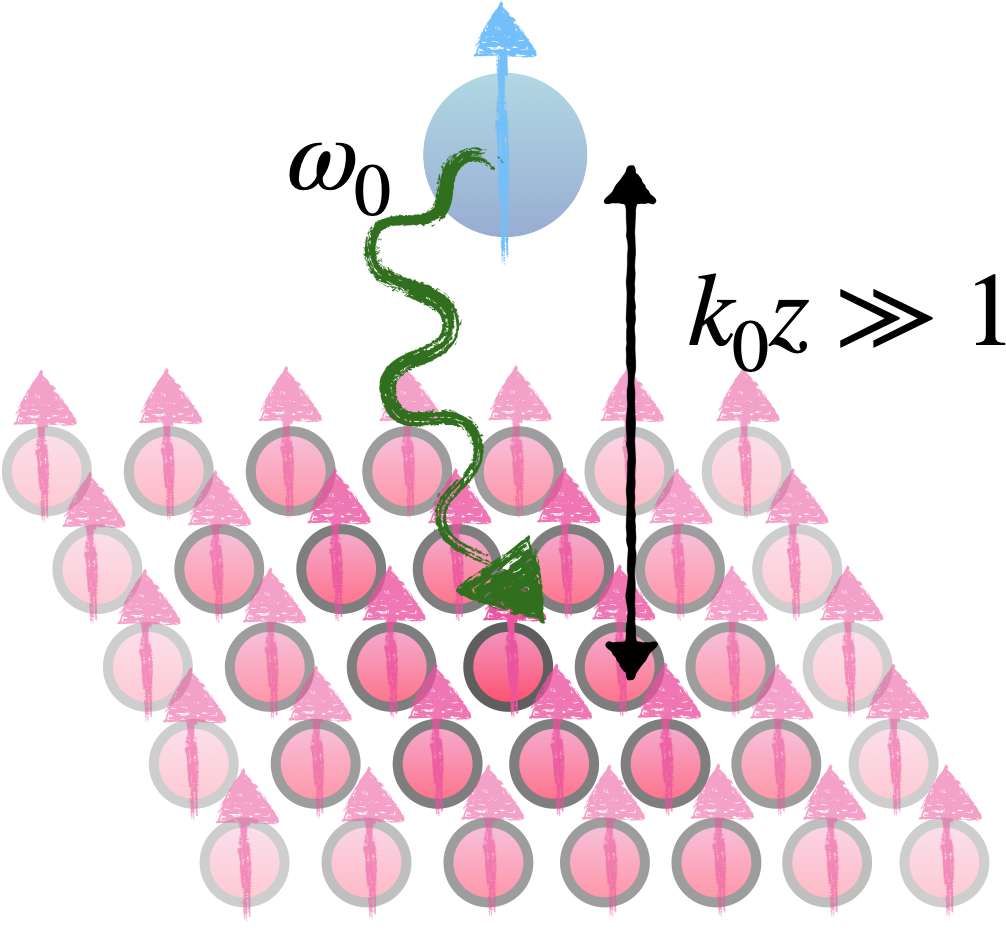}\\Eq.~\eqref{Eq:ResonantRetardedZZ} }}& \raisebox{-3\height}{$1/z^4$~\cite{Milonni2015}} & \raisebox{-3\height}{$1/(z^3 a^2)$ } \\
        \hline
       \raisebox{-0.5\height}{\makecell{\\Retarded $\Delta \omega^\mr{OR} $ \\
\includegraphics[width=0.16\textwidth]{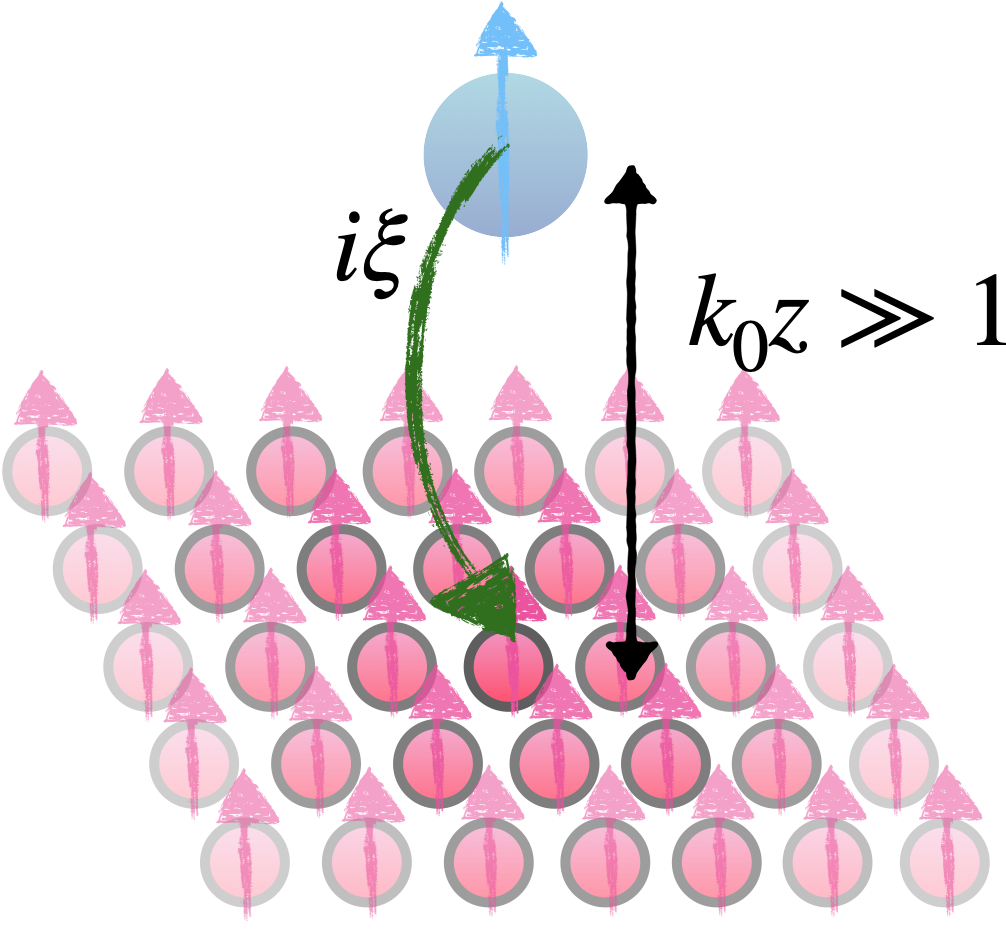}\\Eq.~\eqref{Eq:OfResonantRetardedZZ} }} & \raisebox{-3\height}{$1/z^7$~\cite{CP1948} } & \raisebox{-3\height}{$1/(z^5 a^2)$~\cite{Ye2026}} \\
        \hline
    \end{tabular}
    \caption{ Summary of asymptotic scaling laws for resonant and off-resonant CP potentials seen by the test atom for different lattice spacings and separation between atom and array, assuming all array dipoles are oriented along the $z$-axis. For subwavelength arrays ($ a\ll \lambda_0$) the array behaves as a collective boundary and the CP shift acquires an enhancement factor of  $1/\tilde a^2$ from the areal density of scatterers. In contrast, for $ a\gg\lambda_0$ the test atom is able to resolve individual array atoms, recovering the known two-atom Van der Waals/CP potentials. }
    \vspace{-0.5 cm}\label{Table:CPScalingLawsZZ}
    \end{center}
\end{table}
    \subsubsection{Retarded regime $( z/\lambda_0\gg 1)$}  
    
     The resonant CP shift in the retarded regime is given by (see Appendix~\ref{App:res} for details)
\begin{subequations}\label{Eq:ResonantRetardedZZ}
            \eqn{\label{Eq:ResonantRetardedZZSparse}
\lim_{N\to\infty}\left.\frac{\Delta\omega^{\mr{R}}}{\gamma_0}\right\vert_{\tilde z\gg1;\tilde a\gg1}\approx &\frac{9\gamma_0\omega_M}{2\delta\bkt{\omega_0+\omega_M}}\frac{\cos\bkt{2\tilde z}}{\tilde z^4}\\\label{Eq:ResonantRetardedZZDense}
\lim_{N\to\infty}\left.\frac{\Delta\omega^{\mr{R}}}{\gamma_0}\right\vert_{\tilde z\gg1;\tilde a\ll1}\approx &\frac{9\gamma_0\pi~\omega_M}{2\delta\bkt{\omega_0+\omega_M}}\frac{\sin\bkt{2\tilde z}}{\tilde a^2\tilde z^3}
}
\end{subequations}

     From Eq.~\eqref{Eq:ResonantRetardedZZDense} we see that  in the large lattice-spacing limit ($a\gg \lambda_0$), where the test atom effectively interacts with a single array atom,  the resonant CP potential scales as $\sim1/\tilde z^4$.  In this limit,  $ \Delta \omega^\mr{R}$ agrees  with the CP potential between an  excited and a ground state atom, as obtained in Ref.~\cite{Milonni2015}.  In contrast, for $a\ll \lambda_0$ (Eq.~\eqref{Eq:ResonantRetardedZZSparse}), the resonant CP potential scales as $\sim1/\tilde z^3$.  This is, again, a distinct scaling law that we observe for CP potential in the retarded regime. For comparison, the resonant CP potential near a planar macroscopic medium in the retarded regime scales as $ \sim 1/\tilde z^4$. 
     
     The off-resonant shift in the retarded regime becomes (see Appendix~\ref{App:offres})
    \begin{subequations}
\label{Eq:OfResonantRetardedZZ}
           \eqn{\label{Eq:OffResonantRetardedZZSparse}
\lim_{N\to\infty}\left.\frac{\Delta\omega^\mr{OR}}{\gamma_0}\right\vert_{\tilde z\gg1;\tilde a\gg1}\approx & \frac{45\gamma_0}{8\pi\omega_M}\frac{1}{\tilde z^{7}},\\\label{Eq:OffResonantRetardedZZDense}
\lim_{N\to\infty}\left.\frac{\Delta\omega^\mr{OR}}{\gamma_0}\right\vert_{\tilde z\gg1;\tilde a\ll1}\approx & \frac{9\gamma_0}{10\omega_M}\frac{1}{\tilde a^2\tilde z^5}.
    }
    \end{subequations} 
 
    Eq.~\eqref{Eq:OffResonantRetardedZZSparse} shows that for large lattice spacings the off-resonant contribution to the total CP potential   approaches the conventional $1/\tilde z^7$ dependence of the Casimir-Polder potential between two individual atoms~\cite{CP1948}. For subwavelength lattices ($ a\ll\lambda_0$), the off-resonant CP potential scales as $1/\tilde z^5$. This scaling  agrees with the recent results by Ye et al. \cite{Ye2026} for a ground state isotropic atom near a dense atom array. Since they consider a subwavelength lattice, the single-atom-like regime with $\sim 1/\tilde z^7$ scaling is not observed in their setup.
         
\subsubsection{Resonant CP potential as a function of array size}

         Fig.~\ref{Fig:CPZZPlots} shows  the resonant CP energy shift as a function of $\tilde z$ in  for different values of $N$, obtained via numerically summing over the pairwise CP potential induced by each array atom (Eq.~\eqref{Eq:omegaR}). 
         We observe that regardless of the array size $N$, when the test atom is placed at a separation $ z\ll a$ from the array, the test atom effectively interacts only with one atom of the array with a  scaling of $\sim 1/\tilde z^6$, as captured by the red-dotted line in Fig. \ref{Fig:CPZZPlots} (c), corresponding to Eq. \eqref{Eq:ResonantNonRetardedZZsparse}. As $\tilde z$ increases, the bulk CP contribution becomes prominent with a scaling of $\sim1/\tilde z^4$, captured by the black-dotted line which corresponds to the first term in Eq. \eqref{Eq:ResonantNonRetardedZZdense}.  We see  that as the test atom-array separation $\tilde z $ is of the order of the total length of the array  $\sqrt{N}\tilde a$, the resonant CP shifts deviate significantly from the infinite lattice limit.  Thus for $\tilde z\gtrsim \sqrt N\tilde a$, the  CP shifts are not captured accurately by the expressions for infinite-lattices.  
         
         In the retarded regime, $\tilde z\gg1$, only the $N=10^{10}$ (solid blue) curve mimics the infinite lattice behavior well. By fitting the solid purple line, corresponding to  Eq. \eqref{Eq:ResonantRetardedZZDense}, we deduce that the scaling law of the resonant CP potential in the retarded regime is $\sim 1/\tilde z^3$, in contrast to the typical $ 1/\tilde z^4 $ scaling of the retarded CP potential near a semi-infinite planar  medium~\cite{CP1948, Milonni, Buhmann1}. 

\subsection{Asymptotic CP shifts for $x$-oriented array dipoles}
        \begin{figure}[t]
    \centering
    \includegraphics[width=0.8\linewidth]{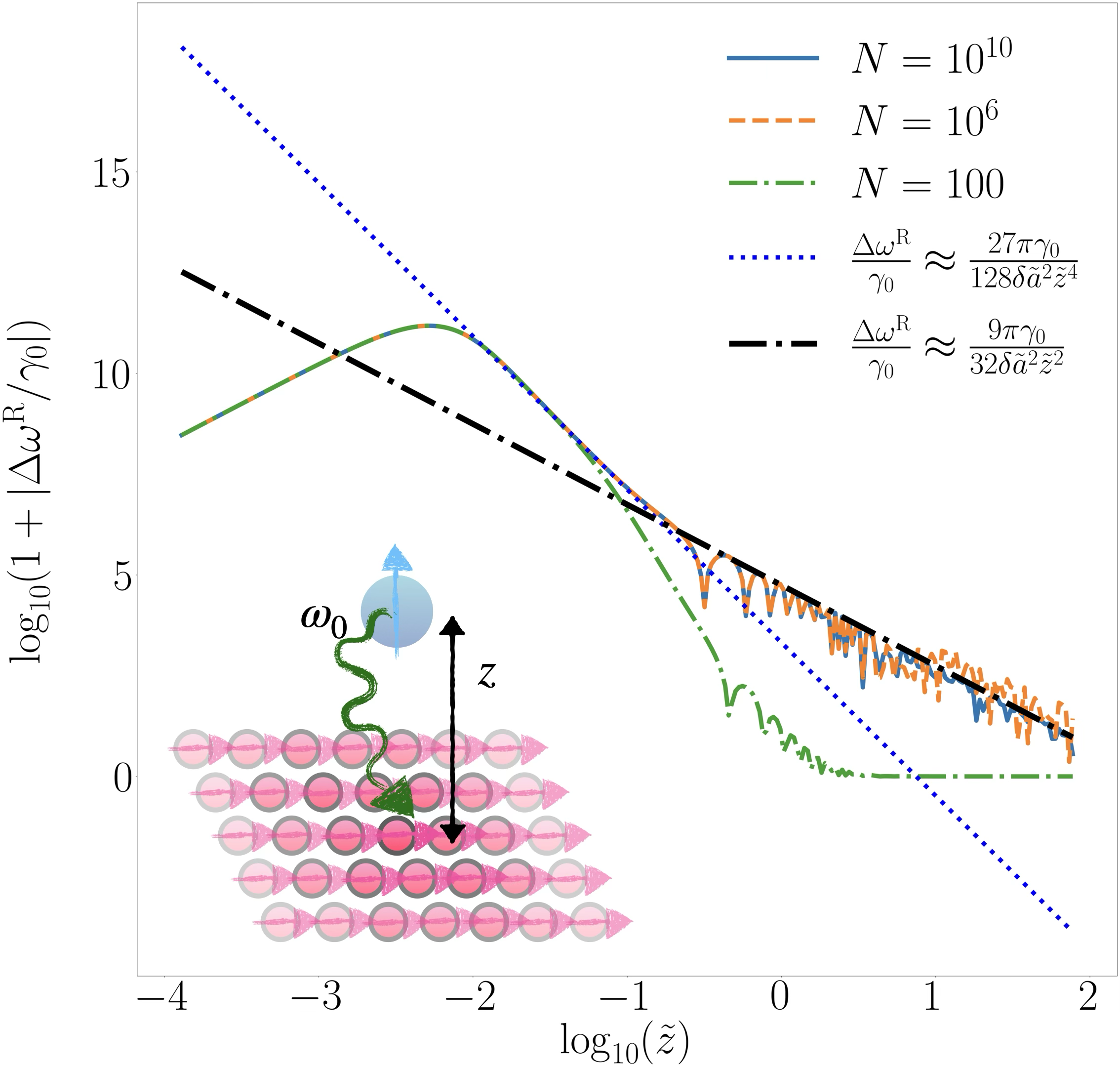}
    \caption{ The resonant shift $\Delta\omega^\mr{R}$ is plotted for $N=100$ (dash-dotted green), $N=10^6$ (dashed orange), and $N=10^{10}$ (solid blue) array atoms, obtained by performing a numerical sum over all lattice points using Eq.~\eqref{Eq:omegaR} for a fixed $\tilde a\approx 10^{-2}$.
    }
    \label{Fig:XZCPPlots}
\end{figure}
We now consider the array atomic dipoles to be oriented along the $x$-axis and analyze the asymptotic CP shifts in different regimes.



    \subsubsection{Non-retarded regime $\bkt{ z\ll\lambda_0 }$} The resonant and off-resonant CP shifts in the non-retarded regime are (see Appendix~\ref{Appendix: Scaling Laws}):
    \eqn{\label{Eq:ResonantZXShiftNonRetarded}
    \left.\frac{\Delta\omega^\mr{R}}{\gamma_0}\right\vert_{\tilde z\ll1; \tilde a\ll1}\approx & \frac{27\gamma_0\pi\omega_M}{32\delta\bkt{\omega_0+\omega_M}}\frac{1}{\tilde a^2\tilde z^4},\\
    \label{EQ:CPORZXNonRetarded}
\lim_{N\to\infty}\left.\frac{\Delta\omega^\mr{OR}}{\gamma_0}\right\vert_{\tilde z\ll1; \tilde a\ll1}\approx& \frac{27\pi \gamma_0}{128\bkt{\omega_0 + \omega_M}}\frac{1}{\tilde a^2\tilde z^4}.
}

We note that for $x$-oriented array dipoles, for sparse lattices ($a \gg \lambda_0$) where the test atom effectively interacts with a single array atom, $U_\mr{CP}\to0$. This agrees  with the known result that for a pair of atoms---one excited and one ground---with orthogonal dipole orientations, the CP potential vanishes~\cite{Milonni2015}. For dense lattices ($ a \ll \lambda_0$), both the resonant and off-resonant CP potentials scale as $\sim1/\tilde z^4$, which is distinct from the typical $ 1/\tilde z^3$ scaling of CP potentials for an atom near a semi-infinite planar medium~\cite{Milonni, CP1948, Buhmann1}. We remark that the $1/\tilde z^4$ scaling of the off-resonant CP potential for a subwavelength array was observed recently in \cite{Ye2026}.  

    \begin{table}[ht]
\begin{center}
     \begin{tabular}{ | c | c| c | } 
    \hline
         \makecell{CP potential\\ $x$-aligned dipoles} & \makecell{Single-atom limit\\ $a\gg\lambda_0,z$}& \makecell{Dense-array limit\\ $a\ll\lambda_0,z$} \\
         \hline
          \raisebox{-0.5\height}{\makecell{\\
Non-retarded $\Delta \omega^\mr{R}$\\
\includegraphics[width=0.16\textwidth]{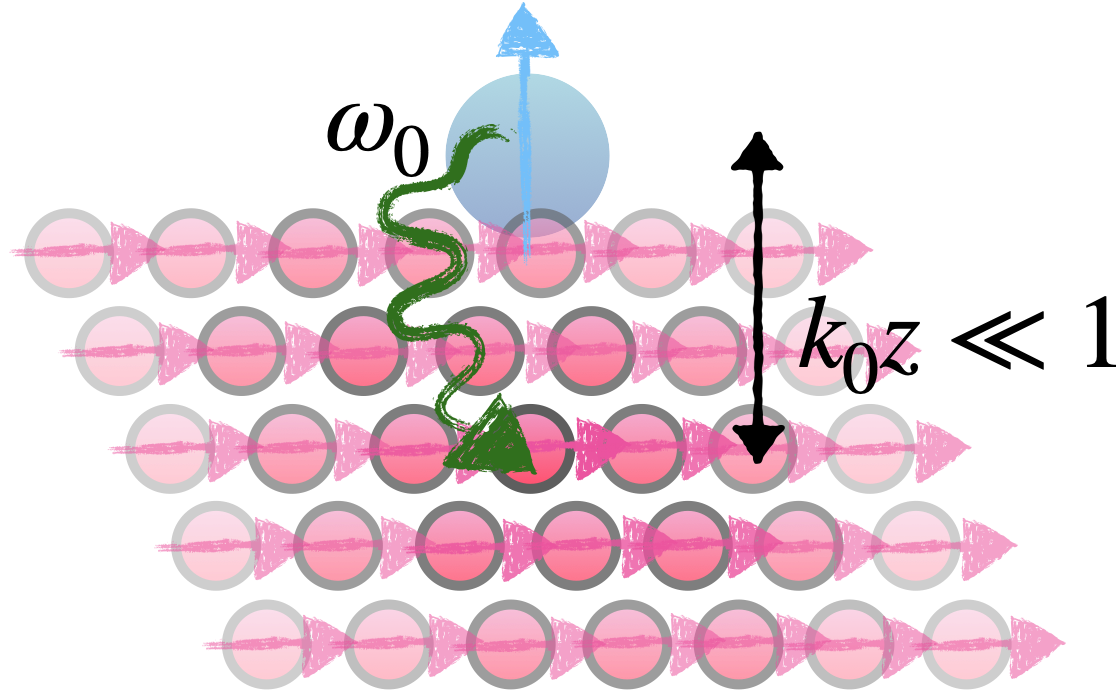}\\
Eq.~\eqref{Eq:ResonantZXShiftNonRetarded}}} & \raisebox{-3\height}{ \makecell{$0$~\cite{Milonni2015}}} & \raisebox{-3\height}{ \makecell{$1/(z^4a^2)$}}  \\
         \hline
          \raisebox{-0.5\height}{\makecell{\\Non-retarded $\Delta \omega^\mr{OR}$\\\includegraphics[width=0.16\textwidth]{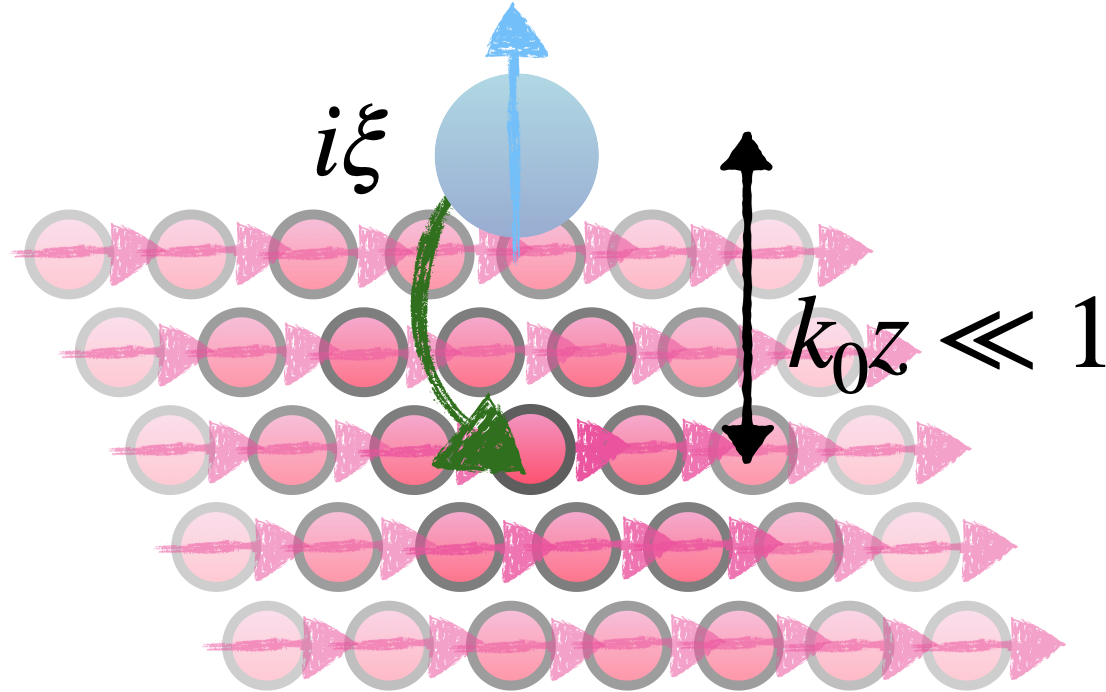}\\
Eq.~\eqref{EQ:CPORZXNonRetarded}}} &\raisebox{-3\height}{ \makecell{$0$}}&   \raisebox{-3\height}{ \makecell{$ 1/(z^4 a^2)$~\cite{Ye2026}}}\\
         \hline
               \raisebox{-0.5\height}{\makecell{\\Retarded $\Delta \omega^\mr{R}$\\\includegraphics[width=0.16\textwidth]{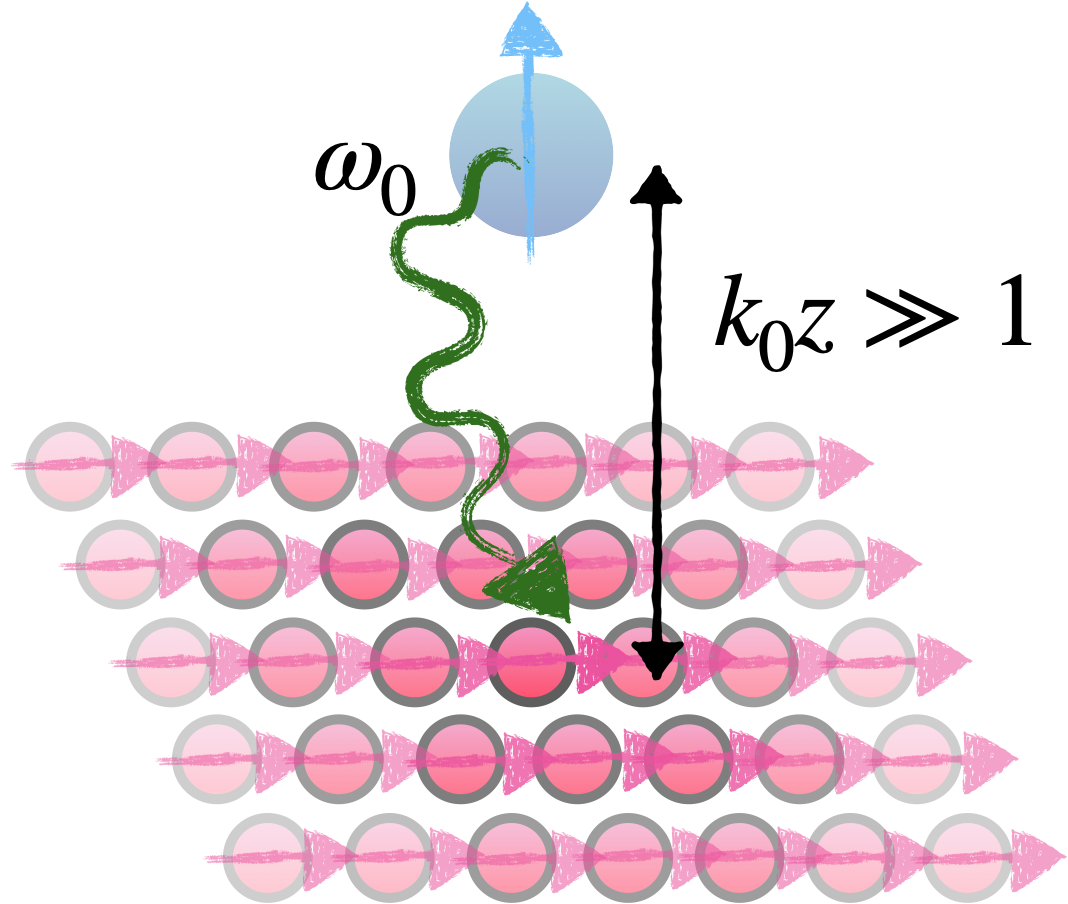}\\
                Eq.~\eqref{Eq:ResonantZXShiftRetarded} }} &\raisebox{-3\height}{$0$ ~\cite{Milonni2015}}& \raisebox{-3\height}{$1/(z^2 a^2)$ }\\
        \hline
      \raisebox{-0.5\height}{\makecell{\\Retarded $\Delta \omega^\mr{OR}$\\\includegraphics[width=0.16\textwidth]{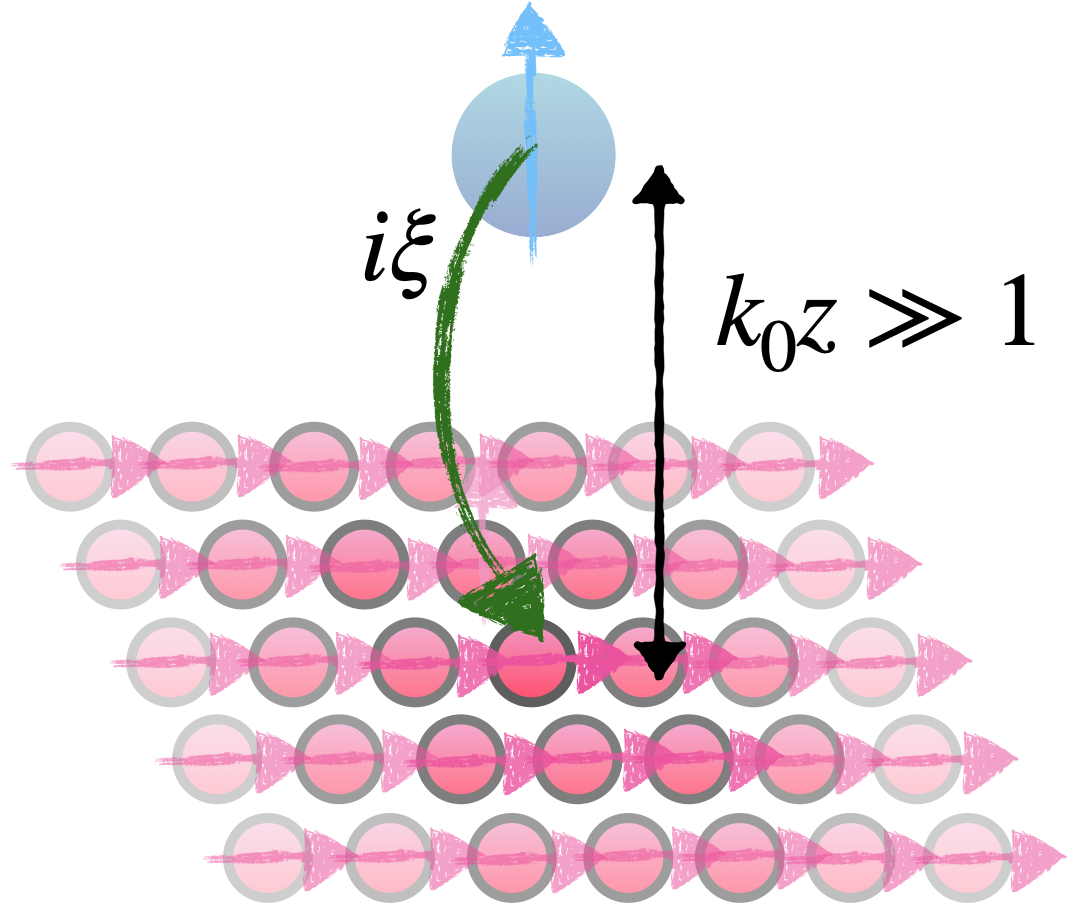}\\
Eq.~\eqref{Eq:RetardedORZXShift}}} &  \raisebox{-3\height}{$0$} & \raisebox{-3\height}{$1/(z^5 a^2)$ \cite{Ye2026}}\\
        \hline
    \end{tabular}
    \caption{Summary of asymptotic scaling laws for resonant and off-resonant CP potentials seen by the test atom for different lattice spacings and separation between atom and array, assuming all array dipoles are oriented along the $x$-axis. For subwavelength arrays ($ a\ll \lambda_0$) the array behaves as a collective boundary and  the CP shift acquires an enhancement factor of  $1/\tilde a^2$ from the areal density of scatterers. In contrast, for $ a\gg\lambda_0$ the test atom is able to resolve individual array atoms, recovering the known two-atom Van der Waals/CP potentials. }
    \vspace{-0.5 cm}\label{Table:CPScalingLawsZX}
    \end{center}
\end{table}

    \subsubsection{Retarded regime $\bkt{ z\gg \lambda_0}$}  The resonant CP energy shift in the retarded regime
    \eqn{\label{Eq:ResonantZXShiftRetarded}
    \left.\frac{\Delta\omega^\mr{R}}{\gamma_0}\right\vert_{\tilde z\gg 1;\tilde a\ll1}\approx -\frac{9\gamma_0\pi\omega_M}{8\delta\bkt{\omega_0+\omega_M}}\frac{\cos\bkt{2\tilde z}}{\tilde a^2\tilde z^2}
    }
    falls off as $1/\tilde z^2$ with pendulations as captured by the $\cos\bkt{2\tilde z}$ term ~\cite{Milonni2015}.  This is, again, a distinct scaling law from the typical $\sim1/\tilde z^4$ scaling of retarded CP shifts near planar half spaces~\cite{CP1948, Milonni, Buhmann1}.

    The off-resonant shift in the retarded regime

        \eqn{\label{Eq:RetardedORZXShift}\lim_{N\to\infty}\left.\frac{\Delta\omega^\mr{OR}}{\gamma_0}\right\vert_{\tilde z\gg1;
        \tilde a\ll1}\approx\frac{9\gamma_0}{16\omega_M}\frac{1}{\tilde a^2\tilde z^5}
        }
        falls off as $1/\tilde z^5$~\cite{Ye2026}. 

\subsubsection{Resonant CP potential as a function of array size}
        In Fig. \ref{Fig:XZCPPlots}, we show the resonant CP potential $\Delta\omega^\mr{R}$ as a function of the test atom-array distance $\tilde z$ for different array sizes $ N$. When $\tilde z\ll \tilde a,$ we see that $\Delta\omega^\mr{R}\approx0$, as in this limit the test atom effectively only interacts with one atom in the array and there is no resultant resonant CP potential from that interaction~\cite{Milonni2015}. As the separation is increased such that $\tilde z\sim\tilde a$, the test atom interacts with the other atoms in the lattice and we see that the resonant CP potential scales as $1/\tilde z^{4}$ (dotted blue line obtained from Eq. \eqref{Eq:ResonantZXShiftNonRetarded}). Similar to the previous case with $z$-oriented array dipoles, here, as $\tilde z\gtrsim \sqrt N\tilde a$, the infinite-lattice scaling law no longer holds as can be seen from the deviating dash-dotted green curve corresponding to $N=100$ lattice.  In  the retarded regime such that $ \tilde z\gg 1$ and $\tilde z\lesssim \sqrt{N}\tilde a$, the resonant CP potential scales as $1/\tilde z^2$ (solid black line, obtained from Eq. \eqref{Eq:ResonantZXShiftRetarded}).

\section{Conclusions and Outlook}
\label{Sec:conclusion}

In this work we developed a framework for describing fluctuation-induced Casimir-Polder (CP) shifts of an excited two-level test atom placed near a 2D atomic array that acts as a tunable atomically-controlled boundary. Given that ordered atomic arrays manifest reflective boundaries whose optical response can be controlled at the level of single atoms, such a system allows one to engineer quantum fluctuation phenomena using the microscopic atomic degrees of freedom in contrast to macroscopic media with fixed optical properties.

We derive analytical expressions for resonant and off-resonant components of the CP potential using fourth-order perturbation theory (Eqs.\eqref{Eq:omegaR} and \eqref{Eq:omegaOR}), under the assumption that the test atom resonance is far detuned from those of the array atoms. The resulting CP potential can be written as a sum over pairwise contributions arising from the interaction between the test atom and the individual array atoms. We analyze the resulting CP potentials for different system parameters, focusing, in particular, on the scaling of the CP potential as a function of the test atom-array separation $ z$ (Section~\ref{Sec:param}).

We show that such a system enables us to bridge the two familiar paradigms of (i) two-atom Van der Waals/Casimir-Polder interactions, recovered when the test atom can resolve individual array atoms in the limit of large lattice spacing $(a\gg\lambda_0)$; and (ii) atom-medium interactions, emerging for subwavelength arrays $(a\ll\lambda_0)$ where the test atom interacts with the array as a whole. Such a crossover is made manifest  via an Euler-Maclaurin decomposition of the lattice sum into a `bulk/medium-like' contribution, proportional to the areal density $1/a^2$ of the scatterers, and a `boundary/single-atom-like' component (Eq.\eqref{Eq:EulerMaclaurinExpansion}). 

We demonstrate that the test atom-array CP potential exhibits distinct asymptotic scalings with $z$, and that these scaling laws can be tuned via the microscopic parameters of the atomic array,  as  summarized in Tables~\ref{Table:CPScalingLawsZZ} and \ref{Table:CPScalingLawsZX}. Specifically, we study the effects of the following parameters:
\begin{itemize}
    \item {Lattice spacing $ a$: Tuning the lattice spacing allows one to cross over from a single-atom-like CP potential for a sparse array ($ a \ll\lambda_0$), to a medium-like response for dense arrays ($ a \gg\lambda_0$). For sparse arrays, where the test atom sees effectively a single array atom, the CP potential follows the known scaling laws between two atoms~\cite{Milonni2015}.  In contrast, for a dense atomic array, the CP potentials (both resonant and off-resonant components) exhibit unconventional scalings with $ z$, in comparison to those for an atom near a planar macroscopic medium (see Table~\ref{Table:CPScalingLawsZZ} and \ref{Table:CPScalingLawsZX}). We further note that the CP potentials exhibit a $ 1/a^2$ scaling for dense arrays, arising from an enhanced effective scattering cross-section of the array.}
    \item {Array dipole orientation: The CP potential depends strongly on the dipole orientation of the array atoms, as illustrated in Tables~\ref{Table:CPScalingLawsZZ} and \ref{Table:CPScalingLawsZX}). Dipole orientation of the array atoms controls which components of the Green tensor contribute, thus modifying how the pairwise contributions add up when summed over all lattice points. }
    \item {Array size $N$: We show that the infinite-lattice asymptotics (Eqs. \eqref{Eq:ResonantNonRetardedZZ}) are  valid in the regime where the test atom-array separation $ z$ is smaller than the size of the array $ \sqrt{N}a$ and finite-size effects set in as $z\gtrsim\sqrt{N}a$.}
    \item{Array detuning: The resonant CP shift is inversely proportional to the  array detuning,  $ \Delta \omega^\mr{R}\sim  1/\delta $, indicating that the resonant contribution can be enhanced for smaller detunings. However, the present analysis is only valid in the limit of large $ \delta$, where multi-scattering processes between the array atoms can be ignored.}
\end{itemize}

Overall, our results establish ordered atomic arrays as a versatile platform for tailoring fluctuation-induced forces with microscopic control, providing a unified description that bridges between discrete two-atom interactions and effective atom-medium interactions. Our results agree with previous limiting cases in appropriate limits, demonstrate new knobs for tuning CP potentials (e.g., lattice constant, dipole orientation and detuning of array atoms, array size), paving the way for several future extensions. 

First, the quantum nature of the atomic boundary motivates one to explore a yet novel mechanism for modifying fluctuation forces via the quantum states of the array atoms. For example,  it was recently demonstrated that a one-dimensional ordered array of multilevel atoms can behave as a `quantum mirror', existing in a coherent superposition of being reflecting and transmitting~\cite{Sinha2025}. The electromagnetic boundary condition imposed by such an atomic array depends on the specific quantum states of its constituent atoms---the  QED phenomena on a nearby test atom are thus inextricably linked with the state of the array atoms. The present work can be extended to include such multilevel atom arrays that exhibit state-dependent boundary conditions. Furthermore adding external drives can enable one to modulate the boundary conditions on fast timescales, with the possibility of probing dynamical Casimir-like effects~\cite{Tobar2026}.

Recent experiments have demonstrated the capacity to create large 2D arrays, trapping thousands of atoms with high fidelity in optical lattices~\cite{Park2022}, optical tweezers~\cite{Tao2024,Pause24,Manetsch2025,Holman2026} or by a combination of both techniques~\cite{Gyger2024}. Remarkably, 2D subwavelength arrays of just $ \sim 1000$ Rydberg atoms can manifest mirrors with reflectances of up to $\approx 0.58(3)$~\cite{Rui2020, Srakaew2023}. Among these results, the experiment by Srakaew et al~\cite{Srakaew2023} demonstrated that the optical response of such arrays can be coherently switched from mirror-like to transparent by manipulating the state of a single ancilla Rydberg atom. The emergent boundary conditions in the presence of such atomic arrays must account for the fact that the macroscopic polarization induced in the atomic medium can exist in coherent superpositions. The consequent QED phenomena (e.g., quantum fluctuation effects and cavity QED), in the presence
of such quantum mirrors \cite{Sinha2025} or quantum metasurfaces \cite{Bekenstein20} are, as yet, largely unexplored. Our results are timely in light of the surge in theoretical and experimental interest in 2D atomic arrays, and motivate future studies to consider vacuum QED effects near such boundaries~\cite{Tobar2026}.
\section{Acknowledgements}
We thank Aaron Bartleson, Clemens Jakubec, Chloe Marzano, Peter Milonni and Sean Raglow  for insightful discussions. This work was supported by the National Science Foundation under Grant No. PHY-2418249, by the  Air Force Office of Scientific Research under Award No. FA9550-25-1-0333, by the John Templeton Foundation under Award No. 63626,  by the U.S. Department of Energy, Office of Science under Grant No. DESC0026059, and by grant NSF PHY-2309135 to the Kavli Institute for Theoretical Physics (KITP).


\appendix
\begin{widetext}
\section{Green tensor}
\label{Appendix:Green}
 The coefficients $ \dbar{G}_{\lambda= e,m}\bkt{\mathbf{r},\mathbf{r'},\omega}$ in Eq.~\eqref{Eq:InteractionHamiltonian} are given by~\cite{Buhmann1}
\eqn{
\dbar G_{e}\bkt{\mathbf{r},\mathbf{r'},\omega}=i\frac{\omega^2}{c^2}\sqrt{\frac{\hbar}{\pi\epsilon}}\dbar G\bkt{\mathbf r,\mathbf r',\omega}\non\\
\dbar G_{m}\bkt{\mathbf{r},\mathbf{r'},\omega}=i\frac{\omega}{c}\sqrt{\frac{\hbar}{\pi\epsilon}}\sbkt{\mathbf{\nabla}'\times\dbar G\bkt{\mathbf r',\mathbf r,\omega}}^\mr{T}
}
with $\dbar {G} \bkt{\mathbf{r},\mathbf{r'},\omega}$ as the free-space Green tensor
\eqn{\label{Eq:Green's Tensor Real Space}
G^{\alpha\beta}(\textbf{r},\textbf{r}',k=\omega/c) = \frac{e^{i kr}}{4\pi r}\sbkt{\bkt{1+\frac{ikr-1}{k^2r^2}}\delta_{\alpha\beta}+\bkt{-1 + \frac{3-3ikr}{k^2r^2}}\frac{r^\alpha r^\beta}{r^2}}
}
with $\alpha,\beta\in\cbkt{x,y,z}$ and $r=\abs{\mb r-\mb{r'}}$.
\section{Fourth-order CP potential }
\label{App:4CP}

We outline the general steps for simplifying the perturbative CP potential in Eq.~\eqref{Eq:FourthOrderPerturbation} as follows. 
    Let us take the first process listed in Table~\ref{Table:dp} as an example, where we get the following four expectation values in Eq.~\eqref{Eq:FourthOrderPerturbation}:
    \eqn{
    \bra{m^0 } \hat H_\mr{int} \ket{\alpha^0} = &\int d^3 r_1 \int d \omega_1 \sum_{\lambda_1}  {\mb e}_1\cdot\dbar{G}^{\dagger}_{\lambda_1}\bkt{\mb {r}_1,\mb r_0,\omega_1}\cdot\mb d_0\\
   \bra{l^0 } \hat H_\mr{int} \ket{m^0} =  &\int d^2 r_2 \int d \omega_2 \sum_{\lambda_2} \mb d_\mb{n}\cdot\dbar{G}_{\lambda_2}\bkt{\mb {r}_2,\mb r_\mb{n},\omega_2}\cdot{\mb e}_2 \\
     \bra{k^0 } \hat H_\mr{int} \ket{l^0} = &\int d^3 r_3 \int d \omega_3 \sum_{\lambda_3} \mb d_0\cdot\dbar{G}_{\lambda_3}\bkt{\mb {r}_3,\mb r_0,\omega_3}\cdot {\mb e}_3\\
    \bra{\alpha^0 } \hat H_\mr{int} \ket{k^0} = &\int d^3 r_4 \int d \omega_4 \sum_{\lambda_4} {\mb e}_4\cdot\dbar{G}^\dagger_{\lambda_4}\bkt{\mb {r}_4,\mb r_\mb{n},\omega_4}\cdot \mb d_\mb{n},
    }
    where $\mb{e}_j$  is the unit vector associated with the annihilation (creation) operators $ \hat{ \mb{f}}^{(\dagger)}_{\lambda_j}(\mb{r}_j, \omega_j) $.
    The inner products create Kronecker-deltas $\delta_{\lambda_1\lambda_2}$, $\delta_{\lambda_3\lambda_4}$,   $\delta_{1,2}$ and $\delta_{3,4}$, and Dirac-delta functions  $\delta\bkt{\mb r_1-\mb{r_2}}$, $\delta\bkt{\mb r_3-\mb{r_4}}$, $\delta (\omega_1 - \omega_2)$ and $\delta (\omega_3 - \omega_4)$ . Integrating over these delta functions reduces the sums and integrals to two-variables: $\omega$ and $\omega'$, $\mb r$ and $\mb r'$, and $\lambda$ and $\lambda'$.
    
    Now  we use the fluctuation-dissipation relation~\cite{Buhmann1}: $\sum_\lambda\int d^3r~\mb d_\mb{n}\cdot\dbar{G}\bkt{\mb r,\mb r_\mb{n},\omega}\cdot\dbar{G}^\dagger\bkt{\mb r,\mb r_0,\omega}\cdot\mb d_0=\frac{\mu_0\hbar\omega^2}{\pi}~\mb d_\mb{n}\cdot \text {Im}\dbar{G}\bkt{\mb r_\mb{n},\mb r_0,\omega}\cdot \mb d_0$ and $\sum_\lambda\int d^3r'~\mb d_0\cdot\dbar{G}\bkt{\mb r',\mb r_0,\omega'}\cdot\dbar{G}^\dagger\bkt{\mb r',\mb r_\mb{n},\omega'}\cdot\mb d_\mb{n}=\frac{\mu_0\hbar{\omega'}^2}{\pi}~\mb d_0\cdot \text {Im}\dbar{G}\bkt{\mb r_0,\mb r_\mb{n},\omega'}\cdot \mb d_\mb{n}$ . The integrals over $\omega$ and $\omega'$ are the only integrals that finally remain. The perturbative shift, after summing over all processes, becomes:
    \eqn{ \label{eq:app ucp}
    U_\mr{CP} = \Delta _\alpha ^{(4)} =\sum_{p = \mr{I}}^{\mr{XII}}  \frac{\mu_0 ^2}{\pi^2\hbar}
    \int d\omega \omega^2\int d\omega' {\omega'}^2\frac{\sbkt{\mb d_0\cdot \text {Im}\dbar{G}\bkt{\mb r_\mb{n},\mb r_0,\omega'}\cdot \mb d_\mb{n}}\sbkt{\mb d_\mb{n}\cdot \text {Im}\dbar{G}\bkt{\mb r_0,\mb r_\mb{n},\omega'}\cdot \mb d_0}}{D_p},
    }
    which corresponds to Eq.\eqref{Eq:FourthOrderShift} in the main text.

\section{Calculating the perturbative shift via Feynman diagrams}\label{Appendix: Feynman Diagrams}
We  compute the CP potentials in Eq.\eqref{Eq:FourthOrderShift} using the Feynman diagrams depicted in Table.~\ref{Table:dp} and noting  that the shift resulting from each process $p = \cbkt{\mr{I},\mr{II}\dots\mr{XII}}$ has the same integral over frequencies $\omega$, $\omega'$  with the distinction being the denominators $ D_p$, as  detailed in Table~\ref{Table:dp}. We consider each $ D_p$, collecting terms to create a partial fraction expansion with $\frac{1}{\omega-\omega'}$ or $\frac{1}{\omega+\omega'}$, closely following the approach in \cite{Buhmann1}. In order to facilitate such a partial fraction expansion,  we combine the following terms:

\eqn{
\frac{1}{D_\mr{III}}+\frac{1}{D_\mr{V}} = &-\frac{1}{\bkt{\omega-\omega_0}\bkt{\omega+\omega_M}\bkt{\omega'-\omega_0}}\\
\frac{1}{D_\mr{VI}}+\frac{1}{D_\mr{VII}}=& -\frac{1}{\bkt{\omega-\omega_0}\bkt{\omega'+\omega_M}\bkt{\omega+\omega_M}}\\
\frac{1}{D_\mr{I}} +\frac{1}{D_\mr{XI}}=&
\frac{1}{\bkt{\omega+\omega'}\delta}\sbkt{\frac{1}{\omega'+\omega_M}+\frac{1}{\omega-\omega_0}}\\
\frac{1}{D_\mr{IX}} + \frac{1}{D_\mr{XII}}=&\frac{1}{\bkt{\omega+\omega'}\delta}\sbkt{\frac{1}{\omega+\omega_M}+\frac{1}{\omega'-\omega_0}}
}

With this, we relabel the denominators as follows:
\eqn{
&\frac{1}{D_1} \equiv \frac{1}{D_\mr{I}} + \frac{1}{D_\mr{XI}} = \frac{1}{\bkt{\omega+\omega'}\delta}\sbkt{\frac{1}{\omega'+\omega_M}+\frac{1}{\omega-\omega_0}}\\
&\frac{1}{D_2} \equiv \frac{1}{D_\mr{II}} =  \frac{1}{\bkt{\omega-\omega_0}\bkt{\omega'-\omega_0}\delta } = \frac{1}{\bkt{\omega-\omega'}\delta}\sbkt{\frac{1}{\omega'-\omega_0}-\frac{1}{\omega-\omega_0}}\\
&\frac{1}{D_3} \equiv \frac{1}{D_\mr{III}} + \frac{1}{D_\mr{V}}  =- \frac{1}{\bkt{\omega-\omega_0}\bkt{\omega+\omega_M}\bkt{\omega'-\omega_0}} = -  
 \frac{1}{(\omega+\omega_M)(\omega -\omega')}\sbkt{\frac{1}{\omega'-\omega_0}-\frac{1}{\omega-\omega_0}}\\
&\frac{1}{D_4} \equiv \frac{1}{D_\mr{IV}} = - \frac{1}{\bkt{\omega-\omega_0}\bkt{\omega+\omega'}\bkt{\omega+\omega_M}}\\
&\frac{1}{D_5} \equiv \frac{1}{D_\mr{VI}} + \frac{1}{D_\mr{VII}}= -\frac{1}{\bkt{\omega-\omega_0}\bkt{\omega'+\omega_M}\bkt{\omega+\omega_M}}=-\frac{1}{\bkt{\omega-\omega_0}( \omega-\omega')}\sbkt{\frac{1}{\omega'+\omega_M}-\frac{1}{\omega+\omega_M}}\\
&\frac{1}{D_6} \equiv \frac{1}{D_\mr{VIII}} = \frac{1}{\bkt{\omega'+\omega_M}\bkt{\omega+\omega_M}{\delta}} = \frac{1}{\bkt{\omega-\omega'} \delta}\sbkt{\frac{1}{\omega'+\omega_M}-\frac{1}{\omega+\omega_M}} \\
&\frac{1}{D_7} \equiv \frac{1}{D_\mr{IX}}+ \frac{1}{D_\mr{XII}}= \frac{1}{\bkt{\omega+\omega'}\delta}\sbkt{\frac{1}{\omega+\omega_M}+\frac{1}{\omega'-\omega_0}}\\
&\frac{1}{D_8} \equiv \frac{1}{D_\mr{X}}= -\frac{1}{\bkt{\omega-\omega_0}\bkt{\omega+\omega'}\bkt{\omega+\omega_M}}
}

Now, by performing the sum of $\frac{1}{D_1}+\frac{1}{D_4}+\frac{1}{D_7}+\frac{1}{D_8} $ we can obtain 
a term of $\frac{1}{\omega+\omega'}$ form: 
\eqn{ \frac{1}{D_1}+\frac{1}{D_4}+\frac{1}{D_7}+\frac{1}{D_8} = & 
\frac{1}{\bkt{\omega+\omega'}\delta }\sbkt{\frac{1}{\omega'+\omega_M}+\frac{1}{\omega-\omega_0}}+\frac{1}{\bkt{\omega+\omega'} \delta }\sbkt{\frac{1}{\omega+\omega_M}-\frac{1}{\omega'-\omega_0}}\non\\
& -\frac{1}{\bkt{\omega-\omega_0}\bkt{\omega+\omega'}\bkt{\omega+\omega_M} }-\frac{1}{\bkt{\omega-\omega_0}\bkt{\omega+\omega'}\bkt{\omega+\omega_M} }
}
Now, the integrals are all symmetric under the exchange $\omega\leftrightarrow\omega'$; we rewrite the above sum as
\eqn{\frac{1}{D_1}+\frac{1}{D_4}+\frac{1}{D_7}+\frac{1}{D_8} = &
\frac{1}{\bkt{\omega+\omega'} \delta}\sbkt{\underbrace{\frac{1}{\omega+\omega_M}}_{\omega\leftrightarrow\omega'}+\frac{1}{\omega-\omega_0}}+\frac{1}{\bkt{\omega+\omega'}\delta }\sbkt{\frac{1}{\omega+\omega_M}+\underbrace{\frac{1}{\omega-\omega_0}}_{\omega\leftrightarrow\omega'}}\non\\
& -\frac{2}{\bkt{\omega-\omega_0}\bkt{\omega+\omega'}\bkt{\omega+\omega_M} }
}
where we have only performed the switch on the indicated terms. This sum is simplified to
\eqn{\label{Eq:omegaplusomega'term}
\frac{1}{D_1}+\frac{1}{D_4}+\frac{1}{D_7}+\frac{1}{D_8} =\frac{4(\omega-\delta)}{\delta\bkt{\omega-\omega_0}\bkt{\omega+\omega_M}\bkt{\omega+\omega'} }.
} 

Adding together the remaining terms we have
\eqn{\frac{1}{D_2} + \frac{1}{D_3}  + \frac{1}{D_5} + \frac{1}{D_6} = & \frac{1}{\bkt{\omega-\omega'} \delta}\sbkt{\underbrace{\frac{1}{\omega'-\omega_0}} _{\omega\leftrightarrow\omega'}-\frac{1}{\omega-\omega_0}} -  
 \frac{1}{(\omega+\omega_M)(\omega -\omega')}\sbkt{\underbrace{\frac{1}{\omega'-\omega_0}}_{\omega\leftrightarrow\omega'}-\frac{1}{\omega-\omega_0}}\non\\
 & -\frac{1}{\bkt{\omega-\omega_0}( \omega-\omega')}\sbkt{\frac{1}{\omega'+\omega_M}-\frac{1}{\omega+\omega_M}}+\frac{1}{\bkt{\omega-\omega'} \delta}\sbkt{\underbrace{\frac{1}{\omega'+\omega_M}} _{\omega\leftrightarrow\omega'}-\frac{1}{\omega+\omega_M}} 
}
Switching $\omega\leftrightarrow\omega'$ in the indicated terms, we get
\eqn{\frac{1}{D_2} + \frac{1}{D_3}  + \frac{1}{D_5} + \frac{1}{D_6} = &- \frac{2}{\bkt{\omega-\omega'}\bkt{\omega-\omega_0} \delta} +  
 \frac{2}{(\omega+\omega_M)(\omega -\omega')\bkt{\omega-\omega_0}}-\frac{2}{\bkt{\omega-\omega'}\bkt{\omega+\omega_M} \delta} \\
\label{Eq:omegaminusomega'term}
=& -\frac{4\bkt{\omega-\delta} }{\delta\bkt{\omega-\omega_0}\bkt{\omega+\omega_M} \bkt{\omega-\omega'}}
}
Now adding Eqs. \eqref{Eq:omegaplusomega'term} and \eqref{Eq:omegaminusomega'term}, Eq.~\eqref{eq:app ucp} becomes:
\eqn{
U_\mr{CP} = \frac{4\mu_0^2}{\hbar\pi^2}\sum_{\mb{n}}\int_0^\infty  d\omega \frac{\omega^2(\omega-\delta)\text{Im}g_{0\mb{n}}(\omega)}{\delta\bkt{\omega-\omega_0}\bkt{\omega+\omega_M} }\int_0^\infty d\omega'\omega'^2 \sbkt{\frac{1}{\omega'+\omega}+\frac{1}{\omega'-\omega}}\text{Im}g_{\mb{n}0}(\omega')
}
To perform the $\omega'$ integral, we split $\text{Im}g_{\mb{n}0}(\omega')=\frac{g_{\mb{n}0}(\omega')-g_{\mb{n}0}^*(\omega')}{2i}$ and use Schwartz reflection principle to write
\eqn{
&\int_0^\infty d\omega' \frac{\omega'^2}{2i}\sbkt{\frac{g_{\mb{n}0}(\omega')}{\omega+\omega'} - \frac{g_{\mb{n}0}(-\omega')}{\omega'+\omega}+\frac{g_{\mb{n}0}(\omega')}{\omega'-\omega}-\frac{g_{\mb{n}0}(-\omega')}{\omega'-\omega} }\non\\
=&\int_0^\infty d\omega' \frac{\omega'^2}{2i}\frac{g_{\mb{n}0}(\omega')}{\omega+\omega'}+\int_{-\infty}^0d\omega'\frac{\omega'^2}{2i}\frac{g_{\mb{n}0}(\omega')}{\omega'-\omega}+\int_0^\infty d\omega' \frac{\omega'^2}{2i}\frac{g_{\mb{n}0}(\omega')}{\omega'-\omega}+\int_{-\infty}^0d\omega'\frac{\omega'^2}{2i}\frac{g_{\mb{n}0}(\omega')}{\omega'+\omega}\non\\
=&\int_{-\infty}^\infty d\omega' \frac{\omega'^2}{2i}\frac{g_{\mb{n}0}(\omega')}{\omega+\omega'}+\int_{-\infty}^\infty d\omega'\frac{\omega'^2}{2i}\frac{g_{\mb{n}0}(\omega')}{\omega'-\omega}
}
Performing a contour integral over $ \omega'$, $ U_\mr{CP}$ becomes
\eqn{
 U_\mr{CP}=\frac{2\mu_0^2}{\hbar\pi}\sum_{\mb{n}}\int_0^\infty  d\omega \frac{\omega^4(\omega-\delta)\text{Im}g_{0\mb{n}}(\omega)}{\delta\bkt{\omega-\omega_0}\bkt{\omega+\omega_M} }\sbkt{g_{\mb{n}0}(\omega)+g_{\mb{n}0}(-\omega)}
}
Again, by splitting  $\text{Im}g_{\mb{n}0}(\omega)=\frac{g_{\mb{n}0}(\omega)-g_{\mb{n}0}^*(\omega)}{2i}$ and using the Schwartz reflection principle we write 
\eqn{
 U_\mr{CP}=& \frac{2\mu_0^2}{\hbar\pi}\sum_{\mb{n}}\int_0^\infty  d\omega \frac{\omega^4(\omega-\delta)}{\delta\bkt{\omega-\omega_0}\bkt{\omega+\omega_M} }\frac{g_{0\mb{n}}(\omega)-g_{0\mb{n}}(-\omega)}{2i}\sbkt{g_{\mb{n}0}(\omega)+g_{\mb{n}0}(-\omega)}\\
=& -\frac{i\mu_0^2}{\hbar \pi}\sum_{\mb{n}}\int_0^\infty d\omega \frac{\omega^4(\omega-\delta)}{\delta\bkt{\omega-\omega_0}\bkt{\omega+\omega_M} }\sbkt{g_{0\mb{n}}(\omega)g_{\mb{n}0}(\omega)-g_{0\mb{n}}(-\omega)g_{\mb{n}0}(-\omega) }\\
=& -\frac{i\mu_0^2}{\hbar\pi}\sum_{\mb{n}}\int_0^\infty d\omega \frac{\omega^4(\omega-\delta)}{\delta\bkt{\omega-\omega_0}\bkt{\omega+\omega_M} }g_{0\mb{n}}(\omega)g_{\mb{n}0}(\omega)+\frac{i\mu_0^2}{\hbar\pi}\sum_{\mb{n}}\int_{-\infty}^0 d\omega \frac{\omega^4(\omega+\delta)}{\delta\bkt{\omega+\omega_0}\bkt{-\omega+\omega_M} }g_{0\mb{n}}(\omega)g_{\mb{n}0}(\omega)
}
After performing a Wick rotation, the only residues result from the emitter's resonance. This is to be expected as the emitter is the only excited atom. Ultimately, we recover Eqs.~\eqref{Eq:omegaR} and \eqref{Eq:omegaOR}.
\section{Euler-Maclaurin formula for 2D lattice sum}
\label{App:EM}
For a  2D square array of atoms, the sum over atom indices in Eqs.~\eqref{Eq:omegaR} and \eqref{Eq:omegaOR}  corresponds to the following form 
\eqn{\label{eq:sumf}
\sum_{\mb n} f\bkt{n_x, n_y;\omega}=\sum_{n_x= - \sqrt{N}/2}^{\sqrt{N}/2}\sum_{n_y=-\sqrt{N}/2}^{\sqrt{N}/2}f\bkt{n_x, n_y;\omega},
}
where 
\eqn{\label{Eq:fnomega}
f(\mb n;\omega)\equiv g_{0\mb n} (\omega)g_{\mb n0}(\omega) =\sbkt{\mb{d}_0\cdot \dbar G\bkt{\mb{r}_0,\sqrt{n_x^2a^2+n_y^2a^2},\omega} \cdot\mb{d}_\mb{n} }\sbkt{\mb{d}_\mb{n}\cdot \dbar G\bkt{\sqrt{n_x^2a^2+n_y^2a^2}, \mb{r}_0, \omega}\cdot \mb{d}_0}
}

In what follows, we derive, in the limit of a large number of atoms $N$, an asymptotic Euler-Maclaurin formula that approximates the above sums in Eq.~\eqref{eq:sumf} in terms of a continuous `bulk' integral and additional `boundary' terms~\cite{stegun}.

 Given the  symmetry of the system, i.e. $ f\bkt{n_x, n_y; \omega} = f\bkt{-n_x, n_y; \omega} = f\bkt{n_x, -n_y; \omega} = f\bkt{-n_x, -n_y; \omega}$, we express the above sum as 
\eqn{\label{eq:f4quad}
\sum_{n_x= - \sqrt{N}/2}^{\sqrt{N}/2}\sum_{n_y=-\sqrt{N}/2}^{\sqrt{N}/2}f\bkt{n_x, n_y; \omega} = 4\sum_{n_x= 0}^{\sqrt{N}/2}\sum_{n_y=0}^{\sqrt{N}/2}f\bkt{n_x, n_y; \omega} 
}

We use the one-dimensional Euler-Maclaurin formula~\cite{stegun} for expanding the sum over $n_y$ as follows
\eqn{
\lim_{N\to\infty}\sum_{n_x=0}^{\sqrt{N}/2}\sum_{n_y=0}^{\sqrt{N}/2}f\bkt{n_x,n_y; \omega}=&\sum_{n_x=0}^{\sqrt{N}/2}\left[\int_{0 }^{\sqrt{N}a/2 }\frac{dy}{a}~f\bkt{n_x,y/a; \omega} + \frac{f\bkt{n_x,0; \omega}+f\bkt{n_x,\sqrt{N}/2; \omega}}{2} \right.\non\\
&\left.+ \sum_{k_y=1}^\infty\frac{\mr B_{2k_y}}{\bkt{2k_y}!}\cbkt{f^{2k_y-1}\bkt{n_x,\sqrt{N}/2; \omega} - f^{2k_y-1}\bkt{n_x,0; \omega}}\right],
}
where we have made the change of variable $n_y\rightarrow y/a$ in the integral. 
Here, $\mr B_{2k_y}$ are the Bernoulli numbers and $f^{2k_y-1}$ refers to the order $2k_y-1$ partial derivative of $f$ with respect to $n_y.$

Now, because we assume that the Green functions are well-behaved and vanish at infinities, we can write
\eqn{
\lim_{N\to\infty}\sum_{n_x=0}^{\sqrt{N}/2}\sum_{n_y=0}^{\sqrt{N}/2}f\bkt{n_x,n_y; \omega}=\sum_{n_x}\left[\int_{0}^{\sqrt{N}a/2}\frac{dy}{a}~f\bkt{n_x,y/a; \omega} + \frac{f\bkt{n_x,0; \omega}}{2} - \sum_{k_y=1}^\infty\frac{\mr B_{2k_y}}{\bkt{2k_y}!}{ f^{2k_y-1}\bkt{n_x,0; \omega}}\right],
}
where we have approximated $ f(n_x, \sqrt{N}/2)\rightarrow0$ for large $N$. Importantly, from Eq.~\eqref{Eq:fnomega} we observe that $ f(n_x, n_y; \omega)$ is an even function under the exchange $n_y\to -n_y$ (or $n_x\to -n_x$). Thus, the odd-order derivative of $f$ with respect to $n_y$ $(n_x)$ is odd under the exchange $n_y\to-n_y~(n_x\to-n_x)$. Thus, odd-order $n_y~(n_x)$ derivatives must be equal to $0$ when $n_y =0~(n_x=0)$. This sets every term in the expansion with a Bernoulli number identically equal to $0$, thus yielding
\eqn{
\lim_{N\to\infty}\sum_{n_x=0}^{\sqrt{N}/2}\sum_{n_y=0}^{\sqrt{N}/2}f\bkt{n_x,n_y; \omega}=\sum_{n_x}\left[\int_{0}^{\sqrt{N}a/2}\frac{dy}{a}~f\bkt{n_x,y/a; \omega} + \frac{f\bkt{n_x,0; \omega}}{2} \right].
}

Similarly, applying the Euler-Maclaurin formula to expand the sum over $ n_x$,  we arrive at Eq.~\eqref{Eq:EulerMaclaurinExpansion}.

\section{CP potentials for different array dipole orientations}\label{Appendix: Scaling Laws}
\subsection{Resonant Casimir-Polder shifts}
\label{App:res}
From Eq. \eqref{Eq:EulerMaclaurinExpansion}, we see that for $N\rightarrow\infty$, the sum over lattice positions can be converted to the sum of three terms: (i) an integral with a `bulk term' with $1/a^2$ dependence, (ii) an integral with an `edge term' with $1/a$ dependence and, (iii) a `vertex term' with $1/a^0$ dependence that captures the interaction of the test atom with the lattice atom vertically below it in the lattice. In what follows, we consider the limits of a dense lattice $a\ll\lambda_0$ where the bulk term is the most significant, and the sparse array limit $a\gg\lambda_0$ where only the vertex term contributes significantly.
\subsubsection{Dense array}\label{App:DenseResonant}
When all the array dipoles are $z$-oriented, we have from Eq.\eqref{Eq:omegaR}
\eqn{
 \Delta\omega^{\mr{R}}= & \frac{2d^4\omega_0^4}{\hbar^2 \epsilon_0^2c^4}\sum_\mb{n}\frac{\omega_M}{\delta\bkt{\omega_0+\omega_M} }\mr{Re}\sbkt{\cbkt{\frac{e^{i k_0r_{0\mb{n}}}}{4\pi r_{0\mb{n}}}\cbkt{\bkt{1+\frac{ik_0r_{0\mb{n}}-1}{k_0^2r_{0\mb{n}}^2}}+\bkt{-1 + \frac{3-3ik_0r_{0\mb{n}}}{k_0^2r_{0\mb{n}}^2}}\frac{z^2}{r_{0\mb{n}}^2}}}^2}\non\\
=&\sum_\mb{n}\frac{9\gamma_0^2\omega_M}{8\delta\bkt{\omega_0+\omega_M}\tilde r_\mb{n}^6}\mr{Re}\sbkt{{{e^{2i \tilde r_\mb{n}}}\cbkt{\bkt{\tilde r_\mb{n}^2+{i\tilde r_\mb{n}-1}}+\bkt{-\tilde r_\mb{n}^2 + {3-3i\tilde  r_\mb{n}}}\frac{\tilde z^2}{\tilde r_\mb{n}^2}}}^2}
}
where we define $\tilde r_\mb{n}\equiv k_0\abs{\mathbf{r}_\mb{n}-\mathbf{ r}_0}=2\pi\abs{\mathbf{r}_\mb{n}-\mathbf{ r}_0}/\lambda_0$ and $ \tilde z = k_0 z$ (note that all the array atoms occupy the $ z = 0$ plane). The exponential phase factor leads to pendulations in the CP potential with increasing distance. 

{For a dense array with large $N\rightarrow\infty$, we evaluate the bulk ($1/\tilde a^2$) term of Eq.\eqref{Eq:EulerMaclaurinExpansion} 
\eqn{
\lim_{N\rightarrow\infty}\frac{\Delta\omega^{\mr{R}}}{\gamma_0^2}
\approx&4\lim_{N\rightarrow\infty}\int _{0}^{\sqrt N\tilde a/2}d\tilde x\int _{0}^{\sqrt N\tilde a/2}d\tilde y\frac{9~\omega_M}{8\tilde a^2\delta\bkt{\omega_0+\omega_M}\tilde r^6}\mr{Re}\sbkt{{{e^{2i \tilde r}}\cbkt{\bkt{\tilde r^2+{i\tilde r-1}}+\bkt{-\tilde r^2 + {3-3i\tilde r}}\frac{\tilde z^2}{\tilde r^2}}}^2},
}
 where $\tilde r = k_0\abs{\mathbf{r}_0-\mathbf{r}}$ is the unitless distance of the test atom from any point of the plane at $z=0$.  Here we switch to polar coordinates such that $\int_{0}^{\sqrt N\tilde a/2}d\tilde x\int_{0}^{\sqrt N\tilde a/2}d\tilde y\to\int_0^{\sqrt {N/2}\tilde a}dR ~R\int_0^{\pi/2}d\phi$ 
\eqn{
\lim_{N\rightarrow\infty}\frac{\Delta\omega^{\mr{R}}}{\gamma_0^2}\approx\lim_{N\rightarrow\infty}\int _{0}^{\sqrt {N/2}\tilde a}dR R\frac{9\pi~\omega_M}{4\tilde a^2\delta\bkt{\omega_0+\omega_M}\tilde r^6}\mr{Re}\sbkt{{{e^{2i \tilde r}}\cbkt{\bkt{\tilde r^2+{i\tilde r-1}}+\bkt{-\tilde r^2 + {3-3i\tilde r}}\frac{\tilde z^2}{\tilde r^2}}}^2}.
}
 Then we make the substitution $\tilde r^2=R^2+\tilde z^2\implies\tilde r~d\tilde r=R~dR$
\eqn{
\lim_{N\rightarrow\infty}\frac{\Delta\omega^{\mr{R}}}{\gamma_0^2}\approx\int _{\tilde z}^{\infty}d\tilde r \frac{9\pi~\omega_M}{4\tilde a^2\delta\bkt{\omega_0+\omega_M}\tilde r^5}\re\sbkt{{{e^{2i \tilde r}}\cbkt{\bkt{\tilde r^2+{i\tilde r-1}}+\bkt{-\tilde r^2 + {3-3i\tilde r}}\frac{\tilde z^2}{\tilde r^2}}}^2}
}
For an infinite lattice we can compute 
\eqn{\label{Eq:zomegardense}
\left.\lim_{N\to\infty}\frac{\Delta\omega^{\mr{R}}}{\gamma_0^2}\right\vert_{\tilde a\ll1}
\approx& \frac{9\pi~\omega_M}{16\tilde a^2\delta\bkt{\omega_0+\omega_M}\tilde z^4}\sbkt{-8{\mathcal{C.I}(2\tilde z)}\tilde z^4+\cos(2\tilde z)\bkt{3-2\tilde z^2}+2\tilde z\bkt{3+2\tilde z^2}\sin(2\tilde z)}
}
 By Taylor expanding the cosine-integral term we obtain Eqs. \eqref{Eq:ResonantNonRetardedZZdense} and \eqref{Eq:ResonantRetardedZZDense} in the main text. 

When the array dipoles are oriented along the x-axis and the emitter is oriented along the z-axis,  we obtain the CP potential using Eq. \ref{Eq:Green's Tensor Real Space} as 
\eqn{
\Delta\omega^{\mr{R}}=\sum_\mb{n}\frac{9\gamma_0^2\omega_M}{8\bkt{\omega_0+\omega_M}\delta\tilde r_\mb{n}^6}\re\sbkt{{{e^{2i \tilde r_\mb{n}}}\cbkt{\bkt{-\tilde r_\mb{n}^2 + {3-3i\tilde r_\mb{n}}}\frac{\tilde z_\mb{n}\tilde x_\mb{n}}{\tilde r_\mb{n}^2}}}^2}
}
Taking the large-$N$ limit, we can express the bulk term in Eq.~\eqref{Eq:EulerMaclaurinExpansion} as
\eqn{\label{Eq:xomegardense}
\lim_{N\rightarrow\infty}\frac{\Delta\omega^{\mr{R}}}{\gamma_0^2}\approx\lim_{N\rightarrow\infty}\int _{0}^{\sqrt {N/2}\tilde a}dR ~R\int_0^{2\pi}d\phi \frac{9~\omega_M}{8\tilde a^2\delta\bkt{\omega_0+\omega_M}\tilde r^6}\re\sbkt{{{e^{2i \tilde r}}\cbkt{\bkt{-\tilde r^2 + {3-3i\tilde r}}\frac{\tilde zR\cos(\phi)}{\tilde r^2}}}^2}\non\\
}
We make the substitution $\tilde r^2=R^2+\tilde z^2\implies\tilde r~d\tilde r=R~dR$ and replace the $R^2=\tilde r^2-\tilde z^2$ to obtain 
\eqn{\label{Eq:ResonantZXShift}
&\left.\lim_{N\to\infty}\frac{\Delta\omega^{\mr{R}}}{\gamma_0} \right\vert_{\tilde a\ll1}\approx \frac{9\gamma_0\pi\omega_M}{32\delta\bkt{\omega_0+\omega_M}\tilde a^2\tilde z^4}\sbkt{\bkt{3-4\tilde z^2}\cos\bkt{2\tilde z}+6\tilde z\sin(2\tilde z)}
}
which reduces to Eqs. \eqref{Eq:ResonantZXShiftNonRetarded} and \eqref{Eq:ResonantZXShiftRetarded} in the appropriate limits.
\subsubsection{Sparse array}\label{App:SparseResonant}
For a sparse array with $\tilde a\gg1$, only the vertex term of Eq.~\eqref{Eq:EulerMaclaurinExpansion} has a significant contribution. When the array dipoles are z-oriented, the vertex term can be evaluated as
\eqn{\label{Eq:omegarsparse}
\left.\lim_{N\rightarrow\infty}\frac{\Delta\omega^\mr{R}}{\gamma_0}\right\vert_{\tilde a\gg1}\approx\frac{9\omega_M\gamma_0}{2\delta\bkt{\omega_0+\omega_M}\tilde z^6}\mr{Re}\bkt{{{e^{2i \tilde z}}\cbkt{1-i\tilde z}}^2}\approx \frac{9\omega_M\gamma_0}{2\delta\bkt{\omega_0+\omega_M}\tilde z^6}\sbkt{ \bkt{1 + \tilde z^2 } \cos\bkt{2 \tilde z } + 2\tilde z \sin \bkt{2\tilde z}}
}
reducing to Eqs. \eqref{Eq:ResonantNonRetardedZZsparse} and \eqref{Eq:ResonantRetardedZZSparse} in the non-retarded and retarded limits. When the array dipoles are x-oriented, the vertex term is $0$ and there is no CP potential in this limit.
 }
\subsection{Off-resonant Casimir-Polder shifts}
\label{App:offres}
\subsubsection{Dense array}
The off-resonant part of the CP potential  given  by Eq.~\eqref{Eq:omegaOR} can be written for $z$-oriented array dipoles as:

\eqn{
\frac{\Delta\omega^\mr{OR}}{\gamma_0^2}\approx \frac{9\omega_M}{8\pi\omega_0^2}\sum_{\mb{n}}\int_0^\infty d\tilde\xi\frac{1}{\bkt{\tilde\xi^2+1}\bkt{\tilde\xi^2+\omega_M^2/\omega_0^2} }\frac{e^{-2\tilde\xi \tilde r_{\mb{n}}}}{\tilde r_\mb{n}^6}\sbkt{{\tilde\xi^2\tilde r_{\mb{n}}^2+\tilde\xi \tilde r_{\mb{n}}+1}-{\cbkt{\tilde\xi^2\tilde r_{\mb{n}}^2+3\tilde\xi\tilde r_{\mb{n}}+3}}\frac{\tilde z^2}{\tilde r_\mb{n}^2} }^2
}
where, $\tilde \xi=\xi/\omega_0$. 

In the retarded regime, $\tilde z\gg1$, the exponential decays fast once $\tilde \xi\sim1/\tilde r_\mb{n}$. Therefore, we can consider that only  small values of $\tilde \xi$ contribute to the integral. In this case, we perform the integral over $ \tilde \xi $ first
\eqn{
 \frac{\Delta\omega^\mr{OR}}{\gamma_0^2}\approx &\frac{9}{8\pi\omega_M}\sum_\mb{n}\int_0^\infty d\tilde\xi~\frac{e^{-2\tilde\xi \tilde r_\mb{n}}}{\tilde r_\mb{n}^6}\sbkt{{\tilde\xi^2\tilde r_\mb{n}^2+\tilde\xi \tilde r_\mb{n}+1}-{\cbkt{\tilde\xi^2\tilde r_\mb{n}^2+3\tilde\xi\tilde r_\mb{n}+3}}\frac{\tilde z^2}{\tilde r_\mb{n}^2} }^2\\
 \approx&\frac{9}{32\pi\omega_M}\sum_\mb{n}\frac{13\tilde r_\mb{n}^4-56\tilde z^2\tilde r_\mb{n}^2+63\tilde z^4}{\tilde r_\mb{n}^{11}}
}
Now using the Euler-Maclaurin formula (Eq.~\eqref{Eq:EulerMaclaurinExpansion}), we evaluate the bulk term as 

\eqn{\label{Eq:ORRetardedZZDense}
&\lim_{N\to\infty}\frac{\Delta\omega^\mr{OR}}{\gamma_0}\approx 4\frac{9\gamma_0}{32\pi\omega_M\tilde a^2}\int_{\tilde z}^\infty d\tilde r\tilde r\int_0^{\pi/2} d\phi \frac{13\tilde r^4-56\tilde z^2\tilde r^2+63\tilde z^4}{4\tilde r^{11}}\\
\implies&\lim_{N\to\infty}\left.\frac{\Delta\omega^\mr{OR}}{\gamma_0}\right\vert_{\tilde z\gg1;\tilde a\ll1}\approx  \frac{9\gamma_0}{10\omega_M\tilde a^2\tilde z^5}.
}
 For the non-retarded limit $\tilde z\ll1$, we first perform the sum over $\mb {n}$ using the Euler-Maclaurin formula before the integral over $ \tilde \xi $. So for $\tilde z\ll1$, the bulk term is
\eqn{
\lim_{N\to\infty}\frac{\Delta\omega^\mr{OR}}{\gamma_0^2}\approx \frac{9\omega_M}{8\pi\omega_0^2\tilde a^2}\int_0^\infty d\tilde\xi\frac{1}{\bkt{\tilde\xi^2+1}\bkt{\tilde\xi^2+\omega_M^2/\omega_0^2} }\int_{\tilde z}^\infty d\tilde r~\tilde r\int_0^{2\pi}d\phi\frac{e^{-2\tilde\xi \tilde r}}{\tilde r^6}\sbkt{{\tilde\xi^2\tilde r^2+\tilde\xi \tilde r+1}-{\cbkt{\tilde\xi^2\tilde r^2+3\tilde\xi\tilde r+3}}\frac{\tilde z^2}{\tilde r^2} }^2\non\\
 \approx \frac{9\omega_M}{32\omega_0^2\tilde a^2}\int_0^\infty d\tilde\xi\frac{1}{\bkt{\tilde\xi^2+1}\bkt{\tilde\xi^2+\omega_M^2/\omega_0^2} } \frac{e^{-2\tilde z\tilde \xi}\sbkt{3+2\tilde z\tilde \xi\cbkt{3 + \tilde z\tilde \xi\bkt{1-2\tilde z\tilde\xi} }-8e^{2\tilde z\tilde\xi}\tilde z^4\tilde\xi^4\cbkt{\mathcal{C.H.I}(2\tilde z\tilde \xi)-\mathcal{S.H.I}(2\tilde z\tilde\xi) } } }{\tilde z^4} 
}
Here, since $\tilde z\ll1$, we keep terms to lowest order in $\tilde z$. We also set the exponent to 1, yielding Eq. \eqref{Eq:OffResonantNonRetardedZZdense}.

We use similar methods to evaluate the off-resonant contributions due to x-oriented array dipoles. In the retarded regime, we start with
\eqn{
 \left.\frac{\Delta\omega^\mr{OR}}{\gamma_0^2}\right\vert_{\tilde z\gg1;\tilde a\ll1}
 \approx\frac{9}{8\pi\omega_M}\sum_{\mb n}\int_0^\infty d\tilde\xi\frac{e^{-2\tilde\xi \tilde r}}{\tilde r^6}\sbkt{{\cbkt{\tilde\xi^2\tilde r^2+3\tilde\xi\tilde r+3}}\frac{\tilde z\tilde x}{\tilde r^2} }^2.
}
After performing the $\tilde \xi$ integral and taking the continuum limit for a large $N$, we arrive at Eq. \eqref{Eq:RetardedORZXShift}.

In the non-retarded regime we first sum over the array dipoles
\eqn{
\left.\frac{\Delta\omega^\mr{OR}}{\gamma_0^2}\right\vert_{\tilde z\ll1;\tilde a\ll1}\approx \frac{9\omega_M}{8\pi\omega_0^2}\int_0^\infty d\tilde\xi\frac{1}{\bkt{\tilde\xi^2+1}\bkt{\tilde\xi^2+\omega_M^2/\omega_0^2} }\sum_{\mb n}\frac{e^{-2\tilde\xi \tilde r}}{\tilde r^6}\sbkt{{\cbkt{\tilde\xi^2\tilde r^2+3\tilde\xi\tilde r+3}}\frac{\tilde z\tilde x}{\tilde r^2} }^2.
}

We convert the sum to an integral with the same method used in Eq.~\eqref{Eq:xomegardense} and perform the integrals over $\tilde r$ and $\phi$ to arrive at
\eqn{
\left.\frac{\Delta\omega^\mr{OR}}{\gamma_0^2}\right\vert_{\tilde z\ll1;\tilde a\ll1}\approx\frac{9\omega_M}{64\omega_0^2\tilde a^2}\int_0^\infty d\tilde\xi\frac{1}{\bkt{\tilde\xi^2+1}\bkt{\tilde\xi^2+\omega_M^2/\omega_0^2} }\sbkt{\frac{e^{-2\tilde \xi\tilde z}\cbkt{3+2\tilde z \tilde\xi (3+2\tilde z\tilde\xi)} }{\tilde z^4} };
}
after performing the $\tilde \xi$ integral we arrive at Eq. \eqref{EQ:CPORZXNonRetarded}. 
\subsubsection{Sparse array}\label{App:SparseArrayOffRes}
For a sparse array with $\tilde a\gg1$, when the array dipoles are z-oriented, the vertex term in Eq.~\eqref{Eq:EulerMaclaurinExpansion} is evaluated in the non-retarded regime as
\eqn{\label{Eq:OffResSparceNonRetZZ}
\lim_{N\rightarrow\infty}\left.\frac{\Delta\omega^\mr{OR}}{\gamma_0}\right\vert_{\tilde z\ll1;\tilde a\gg1}\approx \frac{9\gamma_0\omega_M}{2\pi\omega_0^2}\int_0^\infty d\tilde\xi\frac{1}{\bkt{\tilde\xi^2+1}\bkt{\tilde\xi^2+\omega_M^2/\omega_0^2} }{ {\frac{1}{\tilde z^6}\bkt{1+1\tilde\xi\tilde z}^2} }
}
which yields Eq. \eqref{Eq:OffResonantNonRetardedZZsparse}. In the retarded regime, we get Eq.~\eqref{Eq:OffResonantRetardedZZSparse}.

The vertex terms due $x$-oriented array dipoles vanish in this limit.

\end{widetext}
\bibliography{2DArrayCP}
\end{document}